\documentclass[11pt]{article}

\usepackage{amssymb}
\usepackage{amsmath}
\usepackage{amsthm}
\usepackage{amsbsy}
\usepackage{psfrag}
\usepackage{graphics}
\usepackage{graphicx}
\usepackage{subfigure}
\usepackage{epsfig}
\usepackage{caption}
\usepackage{float}
\usepackage{sidecap}
\usepackage{epsfig}
\usepackage{subfigure}
\usepackage{amsfonts}
\usepackage[T1]{fontenc}

\usepackage{amsmath, amssymb}
\usepackage{latexsym, epsfig}

\newcommand{\Tcmf}{\mbox{\footnotesize $
 {\mbox{\boldmath $T$}} \hspace{-.15cm} \mbox{\raisebox{1.5ex}{$\circ$}}$}}
\newcommand{\Tccf}{\mbox{\footnotesize $
 {\mbox{$T$}} \hspace{-.16cm} \mbox{\raisebox{1.5ex}{$\circ$}}$}}

\newcommand{\Tcm}{\mbox{${\mbox{\boldmath $T$}} \hspace{-.17cm} \mbox{\raisebox{1.5ex}{$\circ$}}$}  }
\newcommand{\Tcmo}{\mbox{$
  {\mbox{\boldmath $T$}}^{(0)} \hspace{-.48cm} \mbox{\raisebox{1.5ex}{$\circ$}}$}  }
  
  \newcommand{\Tcc}{\mbox{${T} \hspace{-.19cm} \mbox{\raisebox{1.5ex}{$\circ$}}$} }
  \newcommand{\Tcf}{\mbox{${T} \hspace{-.16cm} \mbox{\raisebox{1.5ex}{$\circ$}}$} }
 \newcommand{\Tcff}{\mbox{${T} \hspace{-.16cm} \mbox{\raisebox{1.55ex}{$\circ$}}$} }

\newcommand{\Smb}{\mbox{\boldmath $S$}}
\newcommand{\Em}{\mbox{\boldmath $E$}}
\newcommand{\Hm}{\mbox{\boldmath $H$}}

\newcommand{\Lbb}{\mbox{${\mathbb L}$}}

\newcommand{\Abb}{\mbox{${\mathbb A}$}}
\newcommand{\Cbb}{\mbox{${\mathbb C}$}}
\newcommand{\Dbb}{\mbox{${\mathbb D}$}}

\newcommand{\Rm}{\mbox{\boldmath $R$}}
\newcommand{\Sm}{\mbox{\boldmath $S$}}

\newcommand{\me}{\mbox{\boldmath $e$}}

\newcommand{\zero}{\mbox{\boldmath $0$}}

\newcommand{\etam}{ \boldsymbol{\eta}}
\newcommand{\bphi}{\boldsymbol{\Phi}}  
\newcommand{\bth}{ \boldsymbol{\Theta}} 
\newcommand{\bpsi}{ \boldsymbol{\Psi}}

\usepackage{color}

\begin{document}

\newtheorem{thm}{Theorem}[section] \newtheorem{cor}[thm]{Corollary}
\newtheorem{lem}[thm]{Lemma} \newtheorem{prop}[thm]{Proposition} \newtheorem{defn}[thm]{Definition} \newtheorem{rem}[thm]{Remark}

\numberwithin{equation}{section}


\title{Monitoring near-surface depth profile of residual stress in weakly anisotropic media by Rayleigh-wave 
dispersion} 
\author{Yue Chen\footnote{Department of Mathematics and Computer Science, Auburn University at
Montgomery, Montgomery, AL 36117, USA.}, Chi-Sing Man\footnote{Department of Mathematics, University of Kentucky, Lexington, KY 40506-0027, USA.},
Kazumi Tanuma\footnote{Department of Mathematics, Faculty of Science and Technology, Gunma University, Kiryu 376-8515, Japan.}, and Christopher M.\ Kube\footnote{Weapons and Material Research Directorate, U.S.\ Army Research Laboratory, Aberdeen Proving Ground, MD 21005-5069, USA.} }

\date{}
\maketitle





\begin{abstract}
Herein we study the inverse problem on inferring depth profile of near-surface residual stress in a weakly anisotropic medium by boundary measurement of Rayleigh-wave dispersion if all other relevant material parameters of the elastic medium are known. Our solution of this inverse problem is based on a recently developed algorithm by which each term of a high-frequency asymptotic formula for dispersion relations can be computed for Rayleigh waves that propagate in various directions along the free surface of a vertically-inhomogeneous, prestressed, and weakly anisotropic half-space. As a prime example of possible applications we focus on a thick-plate sample of AA 7075-T651 aluminum alloy, which has one face treated by low plasticity burnishing (LPB) that induced a depth-dependent prestress at and immediately beneath the treated surface. We model the sample as a prestressed, weakly-textured orthorhombic aggregate of cubic crystallites and assume that by nondestructive and/or destructive measurements we have ascertained everything about the sample, including the LPB-induced prestress, before it is put into service. Under the supposition that the prestress be partially relaxed but other material parameters remain unchanged after the sample undergoes a period of service, we examine the possibility of inferring the depth profile of the partially relaxed stress by boundary measurement of Rayleigh-wave dispersion.

\begin{flushleft}
{\bf Keywords}: Rayleigh waves, ultrasonic dispersion, stress measurement, acoustoelasticity, textured media, surface conditioning
\end{flushleft}
\end{abstract}

\section{Introduction}
\label{intro}

A common practice to provide lifetime enhancement against fatigue and stress-corrosion cracking of metallic parts (e.g., critical components of aircraft engines, welds in steel structures, etc.) is to impart, through surface-conditioning treatments such as shot peening, sand blasting, laser peening, and low plasticity burnishing, a thin surface layer of compressive residual stress on the parts so treated. The protective compressive stress induced by surface conditioning, however, may relax as a result of thermomechanical loadings experienced by the treated part after it is put into service,
thus compromising the very purpose of the surface-conditioning treatment. To ensure safety and performance, a nondestructive technique should be developed so that retention of the protective compressive stress in the treated parts can be monitored in-situ from time to time, thereby providing a basis for deciding whether a treated part should be taken out of service for replacement or re-conditioning treatment.

The layer of compressive residual stress induced by surface treatment typically starts from the surface and, depending on the specific surface-conditioning technique and processing parameters, runs to a depth that ranges from 0.3 mm to roughly 2 mm. The residual stress $\Tcm$
thus created varies with depth from the surface. At the free surface the principal stress of $\Tcm$
which has the free surface as principal plane is zero.
If another principal stress of $\Tcm$ is plotted against depth
from the surface, the graph typically assumes the shape of a check mark with a long tail (see Fig.\ 2 in Section 5.1):
the principal stress starts negative (i.e., compressive) at the surface and goes through 
a quick dip, then after a blunt turn at a minimum value (i.e., maximum compressive stress) increases monotonically until it becomes slightly tensile and reaches a maximum, and then decays in a long tail to approximately zero while remaining tensile. For
life-prediction purposes, monitoring of not only the surface residual stress
but also the profile and depth of penetration of the protective
stress layer (particularly the maximum compressive principal stresses and their locations) are required, because they all strongly affect the fatigue life and corrosion-crack resistance of the treated part.

The presence of stress in a body affects the velocities of elastic
waves propagating in it. This phenomenon is called the acoustoelastic
effect. There is ample experimental evidence (see, for example,
\cite{MKS, MLLF}) that the presence of a surface layer of inhomogeneous
residual stress in an otherwise homogeneous medium will lead to
the dispersion of Rayleigh waves, the quantitative data of
which can be ascertained by boundary measurements. In this paper we shall explore whether we could monitor the retention of the surface-treatment induced layer of protective compressive stress by measurements of Rayleigh-wave dispersion.

Besides inhomogeneous stress, there are other material characteristics (e.g., inhomogeneity in crystallographic texture, surface roughness) of a treated part that will lead to dispersion of Rayleigh waves, often with effects comparable to or stronger than those due to initial stress (see, e.g., \cite{LV, RN}). Should some such characteristic have also changed after the treated part is put in service, other measurements in addition to Rayleigh-wave dispersion would be needed to infer the depth profile of the stress.
As a first step towards the development of an ultrasonic technique for monitoring stress retention in surface-treated samples, here we will restrict our discussion to the following simple situation: Except for the unknown depth-dependent residual stress, all other relevant material parameters are known. One scenario where this could happen is that we have ascertained all relevant material characteristics of the treated sample, including the residual stress imparted by surface-conditioning, before the sample is put into service.\footnote{In manufacturing practice a large number of samples are produced under virtually the same conditions, and quality-control procedures are in place to ensure that all the samples have nominally the same material characteristics. By wasting some samples if necessary, all the relevant material characteristics of a typical sample can be determined by suitable destructive and/or nondestructive measurements.} 
 After a period of service, the protective residual stress may have suffered from partial or total relaxation, but all other material characteristics of the treated sample remain unchanged after its production. 
 
Under the theoretical framework of linear elasticity with initial stress \cite{Bi, Ho, MC, ML}, Man et al.\ \cite{MNTW} recently presented a general procedure for obtaining a high-frequency asymptotic formula for the dispersion of the phase velocity of Rayleigh waves propagating in a vertically-inhomogeneous, prestressed and anisotropic half-space. As a further development, the general procedure given in \cite{MNTW} was adapted by Tanuma et al.\ \cite{TMC} to the case where the incremental elasticity tensor $\Lbb$ can be written as the sum of an isotropic part $\Cbb^{\rm Iso}$ and a perturbative part $\Abb$.  Under a Cartesian coordinate system where the material medium occupies the half-space $x_3 \leq 0$, the perturbative part $\Abb(\cdot)$, the initial stress $\Tcm(\cdot)$, and the mass density $\rho(\cdot)$ were assumed to be smooth functions of $x_3$. Moreover, the following linearization assumption (*) was made: at the free surface $x_3 = 0$ of the material medium the perturbative part $\Abb(0)$ and  the initial stress $\Tcm(0)$ are sufficiently small as compared with $\Cbb^{\rm Iso}$ that for all expressions and formulas which depend on $\Abb(0)$ and $\Tcm(0)$ it suffices to keep only those terms linear in the components of these tensors. Under this setting, specific formulas are derived \cite{TMC} with which the procedure presented in \cite{MNTW} can be implemented to compute iteratively each term of a high-frequency asymptotic formula for dispersion relations that pertain to Rayleigh waves with various propagation directions. Thus for Rayleigh waves of sufficiently high frequencies, dispersion curves  can be generated by the method developed in \cite{TMC} when requisite data on material and stress are given. Once we have that capability, the inverse problem of inferring stress retention from Rayleigh-wave dispersion can be solved by an iterative approach.

The theory developed in \cite{TMC} is meant for applications that include as typical example ultrasonic measurement of stress in metal structural parts, 
where the perturbative part $\Abb$ in the splitting $\Lbb = \Cbb^{\rm Iso} + \Abb$ of the incremental elasticity tensor is originated from the presence of crystallographic texture and of the prestress $\Tcm$. 
Moreover, the shifts in phase velocities of elastic waves caused by texture and initial stress (with the latter bounded by the yield surface) are typically within 2\% of their values for the corresponding isotropic medium with $\Lbb = \Cbb^{\rm Iso}$, which suggests that linearization assumption (*) 
would be adequate. On the other hand, the theory developed in \cite{TMC} does not take into consideration the effects of surface roughness on Rayleigh-wave dispersion. Several empirical studies (see e.g., \cite{LV, RN}) have shown that if Rayleigh-wave dispersion is used for measurement of stress induced by shot-peening or laser-shock peening, the effect of  surface roughness on the dispersion curves cannot be ignored, for it can totally mask the dispersion due to inhomogeneous stress. Surface conditioning by low plasticity burnishing (LPB), however, is different, for LPB leaves a mirror-smooth finish on processed parts. For the dispersion of Rayleigh waves which have frequencies suitable for interrogation of the compressive stress induced by LPB treatment, surface roughness is not an issue (cf. Figure 17 of \cite{RN}, where dispersion curves of several
IN100 nickel-base superalloy specimens surface-treated by LPB, shot peening, and laser-shock peening are shown in comparison).
In this paper we will study as prime example the possibility of using the high-frequency formula for Rayleigh-wave dispersion developed in \cite{TMC} to infer retention of near-surface compressive stress in a thick-plate sample of an AA 7075-T651 aluminum alloy which was surface-treated by low plasticity burnishing (LPB). 

The plan of this paper is as follows. In Section 2 we present, within the context of linear elasticity with initial stress, the constitutive equation of a prestressed medium which is a polycrystalline aggregate of cubic crystallites that carries an orthorhombic texture. Details on material parameters and texture coefficients specific to the aluminum sample, which serves as the prime example of our present study, are given in Appendix A. In Section 3, after we briefly outline the procedure given in \cite{TMC} to arrive at a high-frequency asymptotic formula for Rayleigh-wave dispersion, we present a theorem and its corollary which will be instrumental for reducing the inverse problem in question to solving systems of linear equations iteratively. Section 4 is devoted to a statement of the inverse problem on monitoring of stress retention and its solution. In Section 5, we apply the theory to a specific inverse problem pertaining to the aluminum sample. There, in Sections 5.2--5.3, we describe how ``experimental'' data on Rayleigh-wave dispersion are simulated over the frequency window from 4 MHz to 70 MHz, with the assumption that ``measured'' phase velocities have accuracy of $\pm 3$\,m/s (i.e., $\pm 0.1$\,\%), where we explain also the rationale behind our choice of frequency window and of the assumed ``measurement'' accuracy.  In Section 5.4, we use the third-order approximation of the dispersion relations to infer the depth profile of the residual stress. In our example the inferred and ``real'' stress profiles match well for the range of depth from 0 to 0.7 mm. In Section 6, we examine the scenario in which experimental conditions (e.g., diffraction errors) rule out the use of data at frequencies lower than 7 MHz. We use the second-order approximation of and the simulated data on the dispersion relations over the frequency window from 7 MHz to 70 MHz to infer the stress profile. Our example shows that the inferred and ``real'' stress profiles match quite well for the range of depth from 0 to 0.5 mm. We end the paper with some closing remarks in Section 7.

\section{Constitutive equation}

The thick-plate sample in question (see \cite{MoM} for details on sample preparation) is that of an AA 7075-T651 aluminum alloy, one face of which was surface-treated by low plasticity burnishing (LPB). The LPB-treatment, in general, would introduce a depth-dependent residual stress, which is compressive at and near the treated surface.
It would also induce changes in material properties (e.g., elastic and acoustoelastic constants) which, in the engineering literature, are qualitatively referred to as the effects of ``cold work". In the theory that we adopt in this paper, such effects are described quantitatively as being caused by changes in surface and near-surface crystallographic texture.

We model the LPB-treated aluminum sample as a prestressed and textured polycrystalline medium, which occupies the half-space $x_3 \leq 0$ under a spatial Cartesian coordinate system $OXYZ$ with $x_3 =0$ being the treated surface whereas the 1- and 2-axis are chosen arbitrarily. We assume that the prestress and all material properties of the polycrystalline medium be macroscopically homogeneous with respect to planar translations for a fixed $x_3$, but they may vary with $x_3$. In what follows, dependence of material tensors, prestress, and texture on $x_3$ will be suppressed except on occasions when we want to emphasize that dependence.

Let the lattice of a fixed single crystal be chosen as reference. The lattice orientation at a sampling point in the polycrystalline medium is specified by a rotation $\Rm$ which brings the reference lattice to the lattice at the sampling point. The crystallographic texture \cite{Bun, M98a, Roe} of a material point in the plane defined by $x_3$ is characterized by an orientation distribution function (ODF) $w(x_3)$ defined on the rotation group SO(3).\footnote{That the ODF is defined on the rotation group is a basic assumption in the classical theory of texture analysis as formulated by Bunge \cite{Bun} and Roe \cite{Roe}. Under this assumption crystallite symmetries are described by subgroups of SO(3), which would nominally exclude the common structural metals (e.g., aluminum, copper, iron, titanium) from consideration. On the other hand, as far as the present study is concerned, the classical theory and a more general theory \cite{DM} with the ODF defined on the orthogonal group O(3) lead effectively to the same constitutive formulas (\ref{C'}) and (\ref{D'}).} Following the convention adopted by Roe \cite{Roe}, we endow SO(3) with volume measure $g = 8\pi^2 \wp_H$, where $\wp_H$ is the Haar measure with $\wp_H(\text{SO(3)}) = 1$. The ODF $w$ can be expressed as an infinite series in terms of the Wigner $D$-functions \cite{Biden, Varsh}:
\begin{equation}  \label{waniso} 
w(\Rm) = \frac{1}{8\pi^2} + \sum_{l=1}^{\infty}\sum_{m= -l}^{l}
\sum_{n= -l}^{l}c^{l}_{mn}D^{l}_{mn}(\Rm). 
\end{equation}
Following Roe, in this paper we work with
\begin{equation} \label{w.coeff}
W_{lmn} = (-1)^{n-m}\sqrt{\frac{2}{2l+1}}\,c^{l}_{mn}
\end{equation}
instead of $c^{l}_{mn}$, and we call them the texture coefficients. If the crystallites in the polycrystal
have no preferred orientations, all the texture coefficients vanish and the ODF reduces to
\begin{equation} \label{wiso}
w = w_{\rm iso} \equiv \frac{1}{8\pi^2}.
\end{equation}

We assume that elastic deformations superimposed on the given polycrystalline medium can be adequately described by the theory of linear elasticity with initial stress \cite{Bi, Ho, ML}. The general constitutive equation in that theory can be written \cite{MC, ML} as 
\begin{equation}
\mbox{\boldmath $S$}=\Tcm +\Hm\,\Tcm+ \Lbb[\mbox{\boldmath $E$}];
\label{constitutive}
\end{equation}
here $\Smb=\bigl( S_{ij}\bigr)$ is the first Piola-Kirchhoff stress, $\Tcm=\bigl(\Tcc_{ij}\bigr)$ the initial
stress, $\Hm=\left({\partial u_i}/{\partial x_j}\right)$ the displacement gradient pertaining to the superimposed
small elastic motion, and $\Em=(\Hm+\Hm^T)/2$ the corresponding infinitesimal strain, where the superscript $T$ denotes transposition; $\Lbb$ is the incremental elasticity tensor which, when regarded as a fourth-order tensor on symmetric tensors, has its components $L_{ijkl}\ (i,j,k,l=1,2,3)$ satisfy the major and minor symmetries. 
Motivated by Hartig's law on the affine dependence of the Young's modulus with strain---an empirical finding supported by ``wholly independent, individual experiments from 1811 to the present ... for one solid after another, including all of the metals" (\cite{Bell}, p.\ 155), for the prestressed and textured polycrystalline medium we regard \cite{M98b, Man}
$\Lbb$ as a function of the ODF $w$ and the initial stress \Tcm, and we write
\begin{equation} \label{L1}
\Lbb[\mbox{\boldmath $E$}]=\Lbb(w, \Tcm)[\mbox{\boldmath $E$}] =\Cbb(w)[\mbox{\boldmath $E$}]+\Dbb(w)[\Tcm,\mbox{\boldmath $E$}],
\end{equation}
where $\Cbb$ is the fourth-order elasticity tensor defined on symmetric tensors $\mbox{\boldmath $E$}$, 
and $\Dbb$ is the sixth-order acoustoelastic tensor defined on ordered pairs of
$\Tcm$ and $\mbox{\boldmath $E$}$, and we replace the functions
$\Cbb(\cdot)$ and $\Dbb(\cdot)$, respectively, by their affine approximation:
\begin{align}
\Cbb &=\Cbb(w)= \Cbb(w_{\rm iso})+ \Cbb'(w_{\rm iso})[w-w_{\rm iso}], \label{c-affine} \\
\Dbb &=\Dbb(w)= \Dbb(w_{\rm iso})+ \Dbb'(w_{\rm iso})[w-w_{\rm iso}], \label{d-affine}
\end{align}
where $\Cbb'(w_{\rm iso})$ and  $\Dbb'(w_{\rm iso})$ denote the Fr\'{e}chet derivative of $\Cbb$ and $\Dbb$ at 
$w=w_{\rm iso}$, respectively. Note that the fourth-order tensor $\Cbb'(w_{\rm iso})[w-w_{\rm iso}]$ and the sixth-order tensor $\Dbb'(w_{\rm iso})[w-w_{\rm iso}]$ depend linearly on the texture coefficients.

When the initial configuration is stress-free and the constituting crystallites have no preferred orientations, i.e.,
$\Tcm = \zero$ and $w = w_{\rm iso}$, the incremental elasticity tensor ${\bf L}$ reduces to the elasticity tensor of classical isotropic elasticity given by
\begin{equation} \label{ciso}
\Cbb(w_{\rm iso})[\mbox{\boldmath $E$}] =\lambda({\rm tr} \mbox{\boldmath $E$}){\bf I}+2\mu \mbox{\boldmath $E$},
\end{equation}
where $\lambda$ and $\mu$ are the Lam\'{e} constants. From (\ref{L1})--(\ref{ciso}), we observe that the incremental elasticity tensor can be expressed as
\begin{equation} \label{L2}
\Lbb = \Cbb^{\rm Iso} + \Abb,
\end{equation}
where $\Cbb^{\rm Iso} = \Cbb(w_{\rm iso})$ and
\begin{equation} \label{A}
\Abb[\cdot] =  \Dbb(w_{\rm iso})[\Tcm, \cdot] + \left(\Cbb'(w_{\rm iso})[w-w_{\rm iso}]\right)[\cdot] + \left(\Dbb'(w_{\rm iso})[w-w_{\rm iso}]\right)[\Tcm, \cdot].
\end{equation}
The isotropic sixth-order tensor $\Dbb(w_{\rm iso})$ is given by the representation formula \cite{M98b}
\begin{equation} \label{diso}
\Dbb(w_{\rm iso})[\Tcm, \Em] = \beta_{1}(\text{tr}\,\Em)(\text{tr}\, \Tcm\,){\bf I}+
\beta_2(\text{tr}\,\Tcm\,)\Em+\beta_3\left((\text{tr}\,\Em)\Tcm+(\text{tr}\,\Em\Tcm\,){\bf I}\right)
 +\beta_4(\Em\Tcm + \Tcm\Em),  
\end{equation}
where $\beta_i$ ($i=1, \cdots, 4$) are material constants.
Aluminum single crystals have cubic crystal symmetry specified by point group $O_h$. As far as the effects of crystallographic texture on the even-order tensors $\left(\Cbb'(w_{\rm iso})[w-w_{\rm iso}]\right)[\cdot]$ and $\left(\Dbb'(w_{\rm iso})[w-w_{\rm iso}]\right)[\Tcm, \cdot]$ are concerned, we may use classical texture analysis and treat \cite{DM} the aluminum crystallites as if their point group were $O$, the proper point group in the same Laue class as $O_h$. 

The surface and near-surface (up to a depth of 0.225 mm) crystallographic texture of the sample at the LPB-treated face were measured by X-ray diffraction and serial sectioning. The texture was found to be essentially constant with depth and was orthorhombic with one of the 2-fold axes of rotational symmetry parallel to $OZ$. In the present paper we simply assume that the sample has a homogeneous texture.  Let $OX'Y'Z'$ be a Cartesian coordinate system which has its coordinate axes parallel to the three 2-fold axes of the orthorhombic texture and the $OZ'$ axis agree with $OZ$. Let $W'_{lmn}$ be the texture coefficients of the sample under the coordinate system $OX'Y'Z'$ and the choice that the reference crystal lattice has its three 4-fold axes parallel to the coordinate axes $OX'$, $OY'$, and $OZ'$. The fourth-order tensor $\Cbb'(w_{\rm iso})[w-w_{\rm iso}]$ and sixth-order tensor $\Dbb'(w_{\rm iso})[w-w_{\rm iso}]$ are then given \cite{M98a, Man} respectively by 
\begin{align}
\left( \Cbb'(w_{\rm iso})[w-w_{\rm iso}]\right)[\Em] &= \alpha \bphi(W'_{400}, W'_{420}, W'_{440})[\Em], \label{C'} \\
\left(\Dbb'(w_{\rm iso})[w-w_{\rm iso}]\right)[\Tcm, \Em] &= \sum_{j=1}^{4} \tilde{b}_j \bpsi^{(i)}(W'_{400}, W'_{420}, W'_{440})[\Tcm, \Em] \nonumber \\
& \hspace{0.5in}+ a \bth(W'_{600}, W'_{620}, W'_{640}, W'_{660})[\Tcm, \Em];  \label{D'}
\end{align}
here $a$ and $\tilde{b}_j$ $(j= 1, \cdots, 4)$ are material constants; $\bphi$ is a fourth-order tensor and $\bpsi^{(j)}\ (j=1,\dots,4)$ are sixth-order tensors defined in terms of the texture coefficients $W'_{400}$, $W'_{420}$, $W'_{440}$, and $\bth$ a sixth-order tensor defined in terms of $W'_{600}$, $W'_{620}$, $W'_{640}$ and $W'_{660}$. The components of these tensors in any $OXYZ$ coordinate system which has the $OZ$-axis agree with the $OZ'$ axis are given explicitly in Appendix A. 

The constitutive equation in question, as defined by (\ref{constitutive}) and (\ref{L2})--(\ref{D'}), has 12 material parameters, namely $\lambda$, $\mu$, $\alpha$, $\beta_i$ $(i = 1, ..., 4)$, $\tilde{b}_j$ $(j = 1, ..., 4)$, and $a$. The values of these material parameters and of the relevant texture coefficients $W'_{lmn}$ that pertain to the aluminum sample in question are given in Appendix A.

\section{Dispersion of Rayleigh waves in weakly anisotropic media with vertically-inhomogeneous initial stress}

In what follows we adopt the following basic assumptions \cite{TMC} on the initial stress $\Tcm = \Tcm(x_3)$, the incremental elasticity tensor $\Lbb = \Lbb(x_3)$, the perturbative part $\Abb$ of $\Lbb$, and the mass density $\rho = \rho(x_3)$ :
\begin{itemize}
\item[(a)] $\Tcm(x_3), \Lbb(x_3)$ and $\rho(x_3)$ are smooth functions\footnote{Here and hereafter we use the term ``smooth function'' to denote an  infinitely differentiable function all of whose derivatives are bounded and continuous.}
of the coordinate $x_3\ (x_3 \le 0)$.
\item[(b)] The initial stress $\Tcm$\ is residual, i.e., it satisfies the equation of equilibrium $\text{div}\,\Tcm = \zero$ for $x_3 < 0$ and the components $\Tcc_{i3}(x_3)\ (i=1,2,3)$  vanish at the surface $x_3 = 0$.
\item[(c)] At the  free surface $x_3=0$,  the perturbative part $\Abb$ of $\Lbb$ and the initial stress $\Tcm$ are sufficiently small as compared with the isotropic part ${\Cbb}^{\rm Iso}$ of $\Lbb$ (i.e., $\|\Tcm(0)\| \ll \|{\Cbb}^{\rm Iso}\|$,  $\|\Abb(0)\| \ll \|{\Cbb}^{\rm Iso}\|$, where $\|\cdot\|$ denotes the Euclidean norm)
that for all expressions and formulas which depend on $\Abb(0)$ and $\Tcm(0)$ it suffices to keep only those terms linear in the components of these tensors.
\end{itemize}
Note that by (\ref{L2}) $\Abb:= \Lbb - {\Cbb}^{\rm Iso}$; hence by assumption (a) the perturbative part $\Abb$ of $\Lbb$ also depends smoothly on $x_3$.

The objective of this paper is to study the inverse problem that pertains to using boundary measurement of Rayleigh-wave dispersion to infer the near-surface depth profile of residual stress induced by surface-conditioning treatments such as low-plasticity burnishing if all other material parameters of the sample are known. Our solution of this inverse problem is based on the algorithm developed in \cite{MNTW,TMC} which, under the aforementioned conditions on $\Tcm$, $\Lbb$, $\Abb$ and $\rho$, can iteratively generate each term in a high-frequency asymptotic formula of the dispersion relation of Rayleigh waves once the propagation direction and all the relevant material parameters including the residual stress depth-profile are specified. In this section we briefly outline the steps (cf.\ \cite{TMC} for details)  to solve the direct problem and present a theorem which will be instrumental to solving the inverse problem.

Consider Rayleigh waves that propagate with phase velocity $v$, wave number $\textsl{k}$, and propagation direction $\etam$ along the traction-free surface of the sample, which is modeled as a vertically-inhomogeneous half-space. Let  ${\bf Z}(v)={\bf Z}(v, \etam, \textsl{k})$ be the $3\times 3$ surface impedance matrix that expresses a linear relationship between the displacements at the free surface and the surface tractions needed to sustain them. In \cite{MNTW, TMC} an algorithm is given by which each term ${\bf{Z}}_i\ (i= 1, 2, 3, \cdots)$ of the asymptotic expansion 
\begin{equation}
 {\bf Z}(v)\, =\, \textsl{k}\,{\bf Z}_0(v)+ {\bf Z}_1(v)+\textsl{k}^{-1}\,{\bf Z}_2(v)+\textsl{k}^{-2}\,{\bf Z}_3(v)+ \cdots 
 \label{Z-asym}
\end{equation}
can be computed iteratively once ${\bf Z}_0$ is determined. Specifically ${\bf Z}_1$ is obtained by solving Lyapunov-type equations (20) and (21) in \cite{TMC}, and ${\bf Z}_m$ $(m= 2, 3, \cdots)$ are obtained by solving equations (23) to (25) in \cite{TMC}. Note that $\textsl{k}{\bf Z}_0$ is the surface impedance matrix of the comparative homogeneous elastic half-space which has its  incremental elasticity tensor, mass density, and initial stress equal to ${\Lbb}(0)$, $\rho(0)$, and $\Tcm(0)$, i.e., their value at the surface $x_3 = 0$, respectively.
Let $\varepsilon := 1/\textsl{k}$. From (\ref{Z-asym}) the truncated sum of the asymptotic expansion for the Rayleigh-wave velocity $v_R$ up to the order $\varepsilon^n$, namely
\begin{eqnarray}
v_R=v_0\ +\ v_1\, \varepsilon\  +\ v_2\, \varepsilon^2\,
+\ \cdots \ +\ v_n\, \varepsilon^n,
\qquad (\varepsilon=1/{\textsl{k}})
\label{nth-v-esp}
\end{eqnarray}
is obtained by applying the implicit function theorem to the approximate secular equation
\begin{eqnarray}
R(v, \varepsilon)=\det\left[\
{\bf Z}_0\ +\ {\bf Z}_1\, \varepsilon\  +\ {\bf Z}_2\, \varepsilon^2\,
+\ \cdots \ +\ {\bf Z}_n\, \varepsilon^n \ \right]=0.
\label{nth-approximatesecular}
\end{eqnarray}
Note that $v_0$ satisfies ${\bf Z}_0(v)=0$.

Under assumption (c), 
we are concerned only with the terms in ${\bf Z}_0(v)$ up to those linear in  $\Tcm(0)$ and $\Abb(0)$,
which leads us to write 
\begin{equation}
{\bf Z}_0(v) \approx {\bf Z}_0^{\rm Iso}(v)+ {\bf Z}_0^{\rm Ptb}(v);
\label{Z_0-expansion}
\end{equation}
here we use the notation $\approx$ to indicate that we are 
retaining terms up to those linear in
  $\Abb(0)$  and  $\Tcm(0)$  and that we are neglecting the higher order terms.
  ${\bf Z}_0^{\rm Iso}(v)$ is of zeroth order in $\Tcm(0)$ and $\Abb(0)$, whereas  ${\bf Z}_0^{\rm Ptb}(v)$ is of first order in $\Tcm(0)$ and $\Abb(0)$. Note that $ \textsl{k}\,{\bf Z}_0^{\rm Iso}(v)$ is the surface impedance matrix pertaining to a homogeneous isotropic elastic half-space  with constitutive equation $\Sm={\Cbb}^{\rm Iso}[\Em]$ and with density $\rho=\rho(0)$.

The following mathematical theorem, which describes how $\Tcm(x_3)$ affects $v_m$ $(m=1, 2, \cdots, n)$ in (\ref{nth-v-esp}), will prove to be instrumental when we study the inverse problem.
\noindent
\begin{thm}
For $m=1,2,\cdots, n$,  $v_m$ in the $m$th-order term of  (\ref{nth-v-esp}) depends on $\Tcm(0)$ and on the  $x_3$-derivatives of $\Tcm(x_3)$ at $x_3=0$ up to those of order $m$; in particular, $v_m$ is of first order in the $m$th-order  $x_3$-derivative of $\Tcm(x_3)$ at $x_3=0$.
\end{thm}
A proof of this theorem is given in Appendix B.

In the inverse problem we will adopt a numerical setting where each component of $\Tcm$ is represented as a polynomial in $x_3$ of degree $n \ (n=1,2,3,\cdots)$\footnote{
From the assumptions (a) and (b)  it follows that $\Tccf_{13}(x_3)=\Tccf_{23}(x_3)=\Tccf_{33}(x_3)=0$. Therefore we will use only the first three equations of (\ref{stress-polynomial}) in the present paper.
On the other hand, when the initial stress $\Tcmf$ is not residual (see for example \cite{TMD}), we will have to keep the last three equations of (\ref{stress-polynomial}).}
\begin{gather}
\Tcc_{11}=\Tcc_{11}(0)+ \sum_{m=1}^n a_m\, {x_3}^{\! m},
\qquad
\Tcc_{22}=\Tcc_{22}(0)+\sum_{m=1}^n b_m\, {x_3}^{\! m},
\qquad
\Tcc_{12}=\Tcc_{12}(0)+\sum_{m=1}^n c_m\, {x_3}^{\! m},
\nonumber 
\\
\Tcc_{13}= \sum_{m=1}^n d_m\, {x_3}^{\! m},
\qquad
\Tcc_{23}=\sum_{m=1}^n e_m\, {x_3}^{\! m},
\qquad
\Tcc_{33}=\sum_{m=1}^n f_m\, {x_3}^{\! m}.
\label{stress-polynomial}
\end{gather}
Here the coefficients $a_m, b_m, c_m, d_m, e_m, f_m\ (m=1,2,\cdots, n)$ are to be determined in the implementation for the inverse problem.

From Theorem 3.1 we immediately obtain

\noindent
\begin{cor}
For $m=1,2,\cdots, n$,  $v_m$ in the $m$th-order term of the expansion (\ref{nth-v-esp}) has the following dependency on the parameters $a_m, b_m, c_m, d_m, e_m, f_m\ (1\le m \le n)$:

\noindent
(1)
$
v_1=v_1(a_1, b_1, c_1, d_1, e_1, f_1)
$ is a first-order function of its arguments.

\noindent
(2)  For $m=2,3,\cdots, n$, the function
$$
v_m=v_m(a_1, b_1, c_1, d_1, e_1, f_1, a_2, b_2, c_2, d_2, e_2, f_2, \cdots, a_m, b_m, c_m, d_m, e_m, f_m)
$$
is of first-order in $a_m, b_m, c_m, d_m, e_m, f_m$.
\end{cor}

\begin{rem}
{\rm Tanuma et al.\ \cite{TMC} provide an algorithm for the computation of the functions $v_m$ $(m= 1, 2, \cdots, n)$ if all relevant material parameters, texture coefficients, and the initial stress $\Tcm$ are available as functions of $x_3$. Conversely, Corollary 3.2 allows us to infer the parameters $a_m, b_m, c_m, d_m, e_m, f_m$ $(m=1, 2, \cdots, n)$ from experimental data on Rayleigh-wave dispersion in six propagation directions as follows. First the parameters $a_1, b_1, \cdots, f_1$ are determined from the six values of $v_1$ for the different propagation directions by solving the six equations on $v_1$, which are linear in the unknowns $a_1, b_1, \cdots, f_1$. Second these values of $a_1, b_1, \cdots, f_1$ are substituted into the function $v_2 = v_2(a_1, b_1, \cdots, f_1, a_2, b_2, \cdots, f_2)$, which is linear in the unknowns $a_2, b_2, \cdots, f_2$. These unknowns are evaluated by solving the six equations on $v_2$ for the six different propagation directions. This iterative process is then repeated for $m = 3, \cdots, n$ to get $a_m, b_m, \cdots, f_m$ from the six equations on $v_m$, which are linear in $a_m, b_m, \cdots, f_m$. A special case where the initial stress $\Tcm$ is residual and $n=3$ is discussed in detail in Section 4.2. \hfill $\Box$}
\end{rem}

\section{An inverse problem on monitoring of stress retention}

The inverse problem in question can be described as follows. Suppose we have ascertained all relevant material characteristics of the treated sample, including the profile of the residual stress imparted by surface-conditioning, before the sample is put into service (say, the sample at state 0). After a service period of the sample, suppose only the residual stress in the sample (say, at state 1) may have changed.  Can we infer the depth profile of the current residual stress from measurement data on the dispersion of Rayleigh waves that propagate in various directions along the free surface of the sample?

Recall that we work with two Cartesian coordinate systems; see Section 2. We choose and fix a spatial Cartesian coordinate system $OXYZ$ such that the material medium occupies the half-space $x_3 \leq 0$ and $x_3 =0$ is the surface treated by low plasticity burnishing (LPB). The fixed 1- and 2-axis are chosen arbitrarily. The $OX'Y'Z'$ system is the material coordinate system which has its coordinate axes parallel to the three 2-fold axes of the orthorhombic texture of the sample and the $OZ'$ axis agrees with the $OZ$ axis. We shall consider Rayleigh waves that propagate in the 2-direction. 
Let $\theta$ be the angle of rotation about the $3$-axis that will bring the $2$-axis to the $2'$-axis. Different propagation directions in the sample are obtained by rotating the material half-space about the 3-axis, i.e., by varying $\theta$. Henceforth we call $\theta$ the propagation direction of the Rayleigh wave (relative to the $2'$-direction of the material half-space).

\begin{figure}[t!]
\centering
\includegraphics[width=0.6\textwidth]{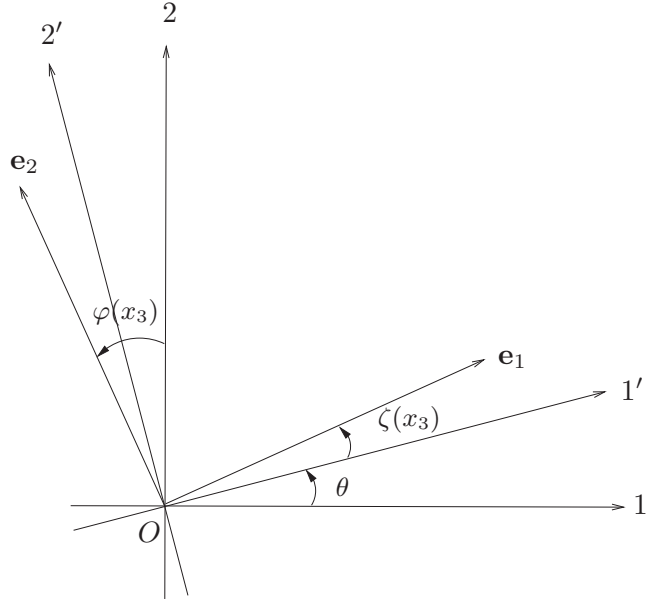}
\caption{Spatial coordinate system, material coordinate system, and principal-stress directions. (Reprinted from \cite{TMC}, with permission from
Elsevier.)}
\label{coordinate}
\end{figure}

Since we assume that the depth-dependent initial stress $\Tcm(x_3)$ be residual, i.e., it satisfies the equation $\text{div}\,\Tcm = {\bf 0}$ for $x_3 < 0$ and the boundary condition of zero traction at $x_3 = 0$, it is of the form
\begin{equation}
\Tcm(x_3)= \left (\begin{array}{ccc}
\Tcc_{11}(x_3) & \Tcc_{12}(x_3) & 0 \\ [1ex]
\Tcc_{12}(x_3) & \Tcc_{22}(x_3) & 0 \\ [1.4ex]
0 & 0 & 0 \end{array} \right )
\label{prestress}
\end{equation}
under the $OXYZ$ coordinate system. 
Let $\mathbf{e}_1(x_3)$ and $\mathbf{e}_2(x_3)$ be the principal directions of the stress that are perpendicular to the 3-axis, and $\sigma_1(x_3)$ and $\sigma_2(x_3)$ be the corresponding principal stresses. Let
$\zeta(x_3)$ be the angle between $\mathbf{e}_2(x_3)$ and the $2'$-axis. Then $\varphi(x_3)=\theta+\zeta(x_3)$ is the angle of rotation about the 3-axis that will bring the direction of the 2-axis to $\mathbf{e}_2(x_3)$; see Fig.\  \ref{coordinate}.
It follows that $\Tcc_{ij}(x_3)$  in (\ref{prestress}) can be written as
\begin{eqnarray}
\Tcc_{11}=\Tcc_m-\Tcc_d\cos{2\varphi},\quad \Tcc_{22}=\Tcc_m+\Tcc_d\cos{2\varphi},\quad
\Tcc_{12}=-\Tcc_d\sin{2\varphi},
\label{rotation}
\end{eqnarray}
where
\begin{eqnarray} \label{Tmd}
\Tcc_m :=\frac{\sigma_1+\sigma_2}{2},\quad \Tcc_d :=\frac{\sigma_2-\sigma_1}{2}.
\end{eqnarray}

Let $\rho_0$ be the density of the aluminum alloy in question when it is stress free. The presence of vertically-inhomogeneous residual stress $\Tcm(x_3)$ will change the density of the material point from $\rho_0$ to $\rho(x_3)$, which is related to $\rho_0$ and $\Tcm(x_3)$ by the formula
\begin{equation} \label{stress-rho}
\rho(x_3)=\rho_0(1-{\rm tr}\Em),\quad \text{ where } \Em =({\Cbb}^{\rm Iso}+ \alpha \bphi)^{-1}[\Tcm].
\end{equation}
In this paper we take $\rho_0 = 2.81 \times 10^3$ kg/m$^3$, which is the (nominal) density of AA7075 alloy as computed from those of its alloying elements and their concentrations (\cite{AA}, pp.\ 2--14).

Now we are ready to make an assertion (**) that serves as an affirmative answer to the question raised at the beginning of this section on the inverse problem:
\vspace*{2ex}

\noindent
(**) Suppose all the relevant material parameters and texture coefficients, except for the residual stress $\Tcm(x_3)$, are known functions of $x_3$. Let $\Tcc_{11}$, $\Tcc_{22}$, and $\Tcc_{12}$ be modeled as polynomials of degree $n$ in $x_3$ as in (\ref{stress-polynomial}). If the parameters $v_0(\theta), v_1(\theta), \cdots, v_n(\theta)$ in the
dispersion relation (cf.\ (\ref{nth-v-esp}))
\begin{equation} \label{disreln}
v_R(\theta) =v_0(\theta) + v_1(\theta)\,\varepsilon  + v_2(\theta)\,\varepsilon^2
+ \cdots + v_n(\theta)\,\varepsilon^n,
\qquad (\varepsilon=1/{\textsl{k}})
\end{equation}
can be evaluated unambiguously from experimental data on Rayleigh-wave dispersion for three different propagation
directions $\theta$, then $\Tcc_{11}(0)$, $\Tcc_{22}(0)$, $\Tcc_{12}(0)$, and all the parameters $(a_m, b_m, c_m)$ for $m= 1, 2, \cdots, n$ in (\ref{stress-polynomial}) can be determined.
\vspace*{2ex}

For definiteness, for the rest of this section we will show how the polynomial model functions $\Tcc_{11}(x_3)$, $\Tcc_{22}(x_3)$, and $\Tcc_{12}(x_3)$ can be determined for the case $n=3$. The dispersion relations in question then reduce to
\begin{equation}
v_R(\theta) = v_0(\theta) +v_1(\theta)\varepsilon+v_2(\theta){\varepsilon}^2+v_3(\theta){\varepsilon}^3,
\label{asymptotic}
\end{equation}
where $0 \leq \theta \leq \pi$, $\varepsilon=\textsl{k}^{-1}$, and $\textsl{k}$ is the wave number. 

\subsection{Determination of $\protect\Tcc_{11}(0)$, $\protect\Tcc_{22}(0)$, and $\protect\Tcc_{12}(0)$}

In (\ref{asymptotic}), $v_0(\theta) = v_0^{\rm Iso}+v_0^{\rm Ptb}(\theta)$ is the zeroth-order term. As
shown by Corollary 8.2 of  \cite{TM}, it is given by the formula
\begin{align}
v_0(\theta) =\ &v_0^{\rm Iso}-\frac{1}{2\rho(0) v_0^{\rm Iso}}
\nonumber\\
& \times \biggl ( A_0+A_2\cos 2\theta +A_4\cos 4\theta +(B_0+B_2\cos 2\theta+B_4\cos 4\theta)\Tcc_m(0)
 \nonumber \\
 & +(C_0+C_2\cos 2\theta+C_4\cos 4\theta+C_6\cos 6\theta)\Tcc_d(0)\cos 2\varphi
\nonumber \\
 &+(D_2\sin 2\theta+D_4\sin 4\theta+D_6\sin 6\theta)\Tcc_d(0)\sin 2\varphi\biggr )
\label{0orderterm}
\end{align}
Here $\rho(0)$ is the density of the material at the free surface $x_3 = 0$. Formulas that express the parameters $A_i\ (i=0,2,4)$, $B_i\ (i=0,2,4)$, $C_i\ (i=0,2,4,6)$, $D_i\ (i=2,4,6)$ and  $v_0^{\rm Iso}$ (the phase velocity of Rayleigh waves in the isotropic base material) in terms of the material parameters and texture coefficients are given in \cite{TM}. Let
\begin{equation}
A:=\Tcc_m(0),\qquad B:=\Tcc_d(0)\cos(2\zeta(0)),\qquad C:=\Tcc_d(0)\sin (2\zeta(0)).
\label{unkowns}
\end{equation}
From (\ref{rotation}) and (\ref{unkowns}),
we get
\begin{align}
\Tcc_{11}(0)&=\Tcc_m(0)-\Tcc_d(0) \cos2(\theta+\zeta)
=A-B \cos 2\theta +C \sin 2\theta, \nonumber
\\[1mm]
\Tcc_{22}(0)&=\Tcc_m(0)+\Tcc_d(0) \cos2(\theta+\zeta)
=A+B \cos 2\theta -C \sin 2\theta,
\\[1mm]
\Tcc_{12}(0)&=-\Tcc_d(0) \sin2(\theta+\zeta)
=-B \sin 2\theta -C \cos 2\theta. \nonumber
\end{align}
Then for a given $\theta$ we can express $\rho(0)$ and thence also $v_0^{\rm Iso}$, through (\ref{stress-rho}), in terms of $A, B, C$ and the known material parameters, texture coefficients, and $\rho_0$. Thus for a given $\theta$ the right-hand side of (\ref{0orderterm}) is a nonlinear function of $A, B$, and $C$, which are the unknowns.  

Let $v_0^{(0)}(\theta)$ and $v_0^{(1)}(\theta)$ be the zeroth-order term for the phase velocities of Rayleigh waves with propagation direction $\theta$ along the free surface of the sample in state 0 and 1, respectively; here the superscript $(j)$ for $j=0,1$ denotes the state 0 or 1. Let
\begin{equation}
\Delta v_0(\theta) =v_0^{(1)}(\theta)-v_0^{(0)}(\theta).
\label{deltav}
\end{equation}
On one hand, $\Delta v_0(\theta)$ in (\ref{deltav}) can be measured by experiments for various $\theta$ at such high frequencies that $v^{(0)}_R$ and $v^{(1)}_R$ appear to become constant. On the other hand, by using the formula (\ref{0orderterm}) to compute $\Delta v_0(\theta)$  we can see that it can be expressed in terms of  $A^{(1)},B^{(1)},C^{(1)}$ which are the unknown parameters (\ref{unkowns}) for state 1.
Thus, we can use Maple to apply the Levenberg-Marquardt method to estimate the parameters $A^{(1)},B^{(1)},C^{(1)}$   with 1 as their initial guess to fit these experimental data. In other words, we can recover $\zeta,\sigma_1,\sigma_2$ at the surface of the material for state 1. From  (\ref{rotation}), we can also evaluate the values of $\Tcc_{11}(0)$, $\Tcc_{22}(0)$, $\Tcc_{12}(0)$ for state 1.  

\subsection{Determination of the parameters $a_m$, $b_m$, and $c_m$ ($m=1,2,3$) in (\ref{stress-polynomial})}

To find the depth profile of the new prestress, we assume that the components of the new prestress can be fitted by some cubic polynomials (cf.\ (\ref{stress-polynomial}))
\begin{align}
\Tcc_{11}(x_3) &=\Tcc_{11}(0)+a_1x_3+a_2x_3^2+a_3x_3^3,\nonumber\\
\Tcc_{22}(x_3) &=\Tcc_{22}(0)+b_1x_3+b_2x_3^2+b_3x_3^3,\label{cubicfit}\\
\Tcc_{12}(x_3) &=\Tcc_{12}(0)+c_1x_3+c_2x_3^2+c_3x_3^3;\nonumber
\end{align}
here $x_3$ denotes the depth;  $\Tcc_{11}(0)$, $\Tcc_{22}(0)$, and $\Tcc_{12}(0)$ have already been determined by the method discussed in the preceding subsection; $a_m, b_m, c_m$ $(m=1, 2, 3)$ are the parameters to be determined. 

To start with, let us choose a specific propagation direction $\theta$. By Corollary 3.2, by applying the algorithm for solving the direct problem, we can get a parametric dispersion relation of the form 
 \begin{align}
 v_R(\theta) &=  v_0(\theta)+v_1(\theta; a_1,b_1,c_1)\,\varepsilon+v_2(\theta; a_1,b_1,c_1,a_2,b_2,c_2)\,\varepsilon^2 \nonumber\\
 &\hspace{0.2in}+v_3(\theta; a_1,b_1,c_1,a_2,b_2,c_2, a_3, b_3, c_3)\,\varepsilon^3,
 \label{pardispersion}
 \end{align}
 where $v_m$ is linear in $(a_m, b_m, c_m)$ for $m = 1,2,3$. Indeed the procedure and formulas for the computations of the functions $v_1$, $v_2$, and $v_3$ are explicitly given in \cite{TMC}. Note that while the values of $v_1(\theta)$, $v_2(\theta)$, and $v_3(\theta)$ in (\ref{asymptotic}) can be determined from experimental data on $v_R(\theta)$ over suitable frequency windows, it is not enough to determine all nine parameters $a_m$, $b_m$, $c_m$ for $m = 1, 2, 3$
 from the system
\begin{gather}
v_1(\theta)=v_1(\theta; a_1,b_1,c_1), \qquad v_2(\theta)=v_2(\theta; a_1,b_1,c_1,a_2,b_2,c_2),\nonumber\\ 
v_3(\theta)=v_3(\theta; a_1,b_1,c_1,a_2,b_2,c_2, a_3, b_3, c_3)
\label{systembi}
\end{gather}
if we use just one $\theta$. Hence we consider three different $\theta$'s to get a complete system.  To determine the nine parameters $a_m, b_m, c_m$, first we solve for $a_1, b_1, c_1$ from the system for $v_1$, which is linear with respect to $a_1, b_1, c_1$. Then substituting the values of $a_1, b_1, c_1$ into the system for $v_2$ to get a linear system with respect to $a_2, b_2, c_2$, we can solve for  $a_2, b_2, c_2$ quickly. After that, continue to substitute the values of $a_1, b_1, c_1, a_2,  b_2, c_2$ into the system for $v_3$. Doing so leads to a linear system associated with $a_3, b_3, c_3$, from which the parameters $a_3, b_3, c_3$ can be easily determined.

 To get better estimates of the parameters $a_m, b_m, c_m$, we can consider several different groups of $\theta$ with each group containing 3 different $\theta$'s. After determining the parameters $a_m, b_m, c_m$ for each group,  we take the average of the parameters from these groups as our estimated values of the parameters $a_m, b_m, c_m$ for $m =1,2,3$.
 
 \section{Recovery of near-surface depth profile of residual stress}
\label{testsample}

\subsection{The ``unknown'' $\protect\Tcm(x_3)$}

\begin{figure}[h!]
\begin{center}
\includegraphics[width=0.5\textwidth]{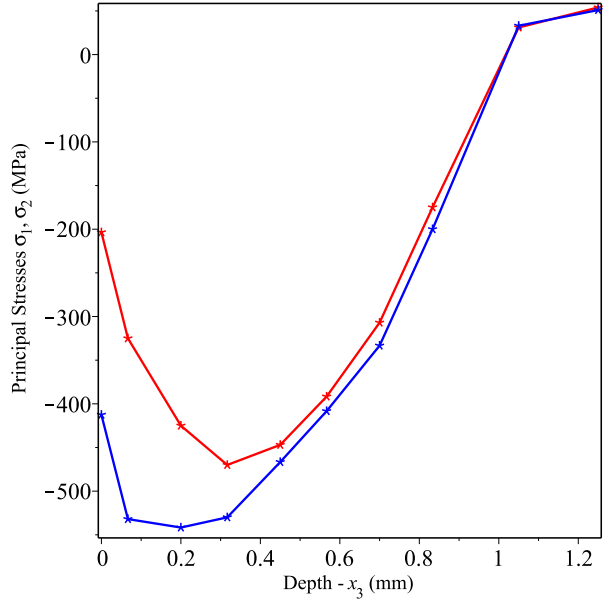}
\caption{Depth profiles of principal stresses $\sigma_1$ (top curve) and $\sigma_2$ (bottom curve)  in sample at state 0. (Reprinted from \cite{TMC}, with permission from 
Elsevier.)} \label{state0}
\end{center}
\end{figure}

The residual stress $\Tcmo$\hspace*{0.4cm} induced by LPB treatment on the AA 7075-T651 sample (state 0) was measured by X-ray diffraction (and supplemented by information gathered from hole-drilling) up to a depth of 1.25 mm from the treated surface.  The depth profiles of the principal stresses are depicted in Figure \ref{state0}, where the top and bottom curves pertain to the principal stresses $\sigma_1(x_3)$ and $\sigma_2(x_3)$, respectively. Since $\Tcmo$\hspace*{0.4cm}$(x_3)$ is also of the form (\ref{prestress}) under the $OX'Y'Z'$ material coordinate system, the principal directions $\me_1(x_3)$ and $\me_2(x_3)$ of the stress are defined with respect to the material coordinate system as soon as the angle $\zeta(x_3)$ between $\me_2(x_3)$ and the $2'$-direction is specified. In the stress measurements it was found that $\zeta(x_3) \approx 10^\circ$ for $0 \geq x_3 \geq -0.5$ mm. As shown in Fig.\  \ref{state0}, $\sigma_1(x_3) \approx \sigma_2(x_3)$ for $x_3 \leq -0.5$ mm. Hence we may  take $\zeta(x_3) \approx 10^\circ$ for $x_3 \leq -0.5$ mm, as $\zeta(x_3)$ is, to within experimental error, arbitrary there. In our computations below, however, only the information on $\Tcmo$\hspace*{0.4cm}$(0)$ will be used to calculate $v^{(0)}_0(\theta)$.

\begin{figure}[h!]
\begin{center}
\includegraphics[width=0.5\textwidth]{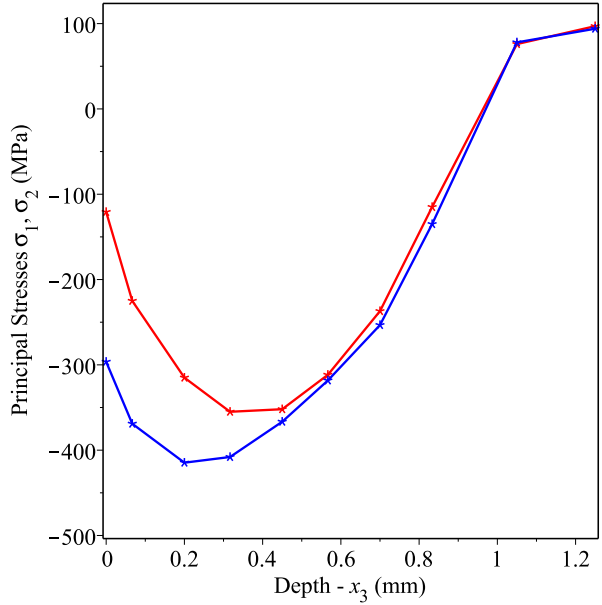}
\caption{Depth profiles of relaxed principal stresses $\sigma_1$ (top curve) and $\sigma_2$ (bottom curve)  in sample at state 1.}
\label{state1}
\end{center}
\end{figure}

Suppose the residual stress has changed to some unknown $\Tcm(x_3)$ (state 1) after the sample is put into service for a period of time, but crystallographic texture and other material parameters of the sample remain the same as before. As an exercise to see if we could determine the unknown $\Tcm(x_3)$ by boundary measurement of Rayleigh-wave dispersion, let us consider a specific instance where the original residual stress is relaxed so that the depth profile of the principal stresses $\sigma_1$ and $\sigma_2$ become those given in Figure \ref{state1}. Moreover the angle $\zeta$ varies with depth as shown in Table 1.

\begin{table}[h!]
\begin{center}
\begin{tabular}{|c|c|c|c|c|c|c|c|c|c|c|}
  \hline
   Depth (mm) & 0 & 0.0667 & 0.2 & 0.3167 & 0.45& 0.5667 &0.7& 0.8333 & 1.05 & 1.25\\
  \hline
   $\zeta$ (degree) &$30^\circ$ & $35^\circ$ & $40^\circ$ & $34^\circ$ & $25^\circ$ & $23^\circ$ & $21^\circ$ & $18^\circ$ & $15^\circ$ & $12^\circ$\\
   \hline
\end{tabular}
\end{center}
\caption{Change of $\zeta$ with respect to depth.}
\label{Table1}
\end{table}
 
\subsection{Experimental considerations}

In this paper we shall test our algorithm for monitoring depth-profile of near-surface residual stress against simulated dispersion data.
To begin we should survey the experimental literature and decide, for our simulated data, what assumptions we will make as regards measurement error of Rayleigh-wave velocity and the frequency window within which dispersion data is available.
Because of the smallness of the acoustoelastic effect, measurement accuracy of Rayleigh-wave velocity is of prime importance to our present study. In this regard, two methods of Rayleigh-wave measurements, namely that which involves scanning acoustic microscopy (SAM) with immersion transducers \cite{Bri} and that which uses laser ultrasonics \cite{Lomo01}, merit special attention.

Rayleigh-wave velocity measurements of stress using SAM systems are well documented \cite{DW96,LK94,MP89}. 
An absolute measurement accuracy of surface-wave velocity of $\pm0.02\%$ has been obtained using a well-controlled and calibrated immersion SAM system \cite{KA98}. To reach this level of accuracy, tedious measures are needed to calibrate for and minimize effects caused by surface abnormalities, temperature fluctuations, and electrical instabilities.  The present study is meant as a first step with the long-term goal to develop
a robust nondestructive evaluation method for measuring depth-dependent residual stress in surface-treated metals, LPB treatments in particular. With this application in mind, it is desirable to perform measurements on \textit{as-received} samples. Whereas the high-frequency SAM systems are highly-sensitive to surface irregularity, laser ultrasonic methods for Rayleigh-wave measurements have demonstrated a balance between surface condition sensitivity and measurement accuracy \cite{RN02,RN,RN04}. Additionally, the flat spectral response common to optical detection systems is appropriate to the stress-dependent dispersion present in the model predictions. With this in mind, the simulations and analysis in this article focus around the capabilities of 
laser-ultrasonic systems.

Using an optical detection system, Ruiz and Nagy \cite{RN04} reported a relative error of $\pm0.1\%$ in the measured Rayleigh-wave velocity. It is this relative error that is chosen to generate the synthetic experimental Rayleigh-wave dispersion data presented in Section 5.3. This level of error is believed to be appropriate and fairly conservative because it was established from measurements performed on rough, shot-peened surfaces with considerable cold-work \cite{RN04}. Recent Rayleigh-wave velocity measurements conducted on the smooth surface of a steel sample under tension were resolved to within $\pm0.005$\% relative to the initial stress-free measurement (cf.\ Figure 4 of \cite{ZL17}). However, achieving measurement error near $\pm0.1\%$ still requires careful experimental considerations.

In general, significant sources of dispersion influencing Rayleigh waves include diffraction, induced surface roughness by the surface treatment, generated residual stresses, and secondary microstructural changes associated with the initial cold work and subsequent thermomechanical loading history \cite{RN04}. For the present consideration, as a first step, we investigate the possibility of monitoring the depth-profile of stress by measurement of Rayleigh-wave dispersion only for situations where crystallographic texture and other material parameters remain unchanged. Moreover, we assume that the mirror-like surface finish produced by the LPB treatment renders the influence of surface roughness on the overall dispersion negligible. However, the smooth surfaces resulting from LPB do not completely eliminate diffraction effects \cite{RN02}. Diffraction is a result of the Rayleigh-wave being generated by a finite-sized source, which causes the wave to exhibit a spatially-dependent phase that leads to self-interaction. Diffraction effects are lessened for detection far from the source or when the Rayleigh wavelength is much smaller than the size of the source, i.e., at high-frequencies. Ruiz and Nagy \cite{RN02} reported that diffraction led to a Rayleigh-wave dispersion on the order of 0.1\% for nominally smooth surfaces. They provided a model-based diffraction correction for Rayleigh-wave dispersion measurements \cite{RN02}. The diffraction correction demonstrated strong agreement with experimental measurements above approximately 3 MHz. Another consideration is that the algorithm we use to solve the inverse problem is based on the high-frequency asymptotic formula (\ref{nth-v-esp}) for Rayleigh-wave dispersion, the details of which are given in \cite{TMC}. The accuracy of the formula depends on the order of approximation and the frequency window we choose. With reference to the comparison of the first-, second-, and third-order terms in the high-frequency asymptotic formula for the aluminum sample in question (see Table 2 of 
\cite{TMC}), we use the third-order approximation and a frequency window from 4 MHz to 70 MHz in Sections 5.3--5.4 and the second-order approximation and a frequency window from 7 MHz to 70 MHz in Section 6.

\subsection{Simulation of velocity and dispersion data}

We assume that measurements of  Rayleigh-wave velocities be made by rotating the sample about the 3-axis, in steps of $15^\circ$, from $\theta=0^\circ$ to $180^\circ$. However, since we have no experimental results in hand, we simulate the velocity and dispersion data of the sample at state 1 as follows.

\begin{figure}[h!]
\begin{center}
\includegraphics[width=0.5\textwidth]{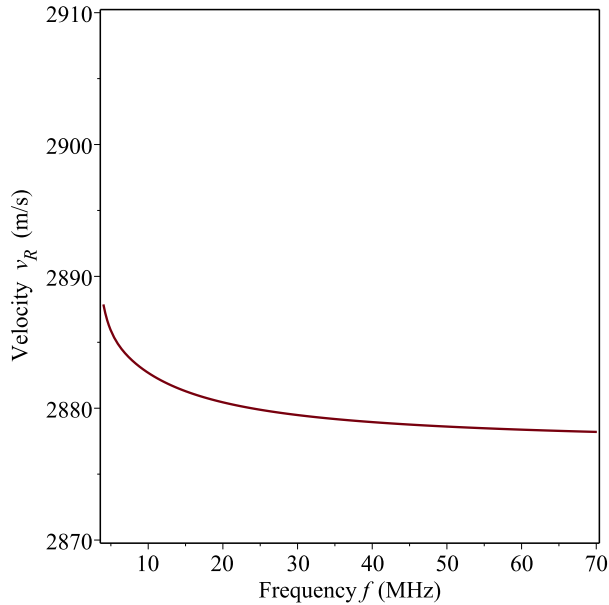}
\caption{Theoretical dispersion curve of sample at state 1 for $\theta=90^\circ$.}
\label{truedis}
\end{center}
\end{figure}

\begin{figure}[t!]
\begin{center}
\includegraphics[width=0.5\textwidth]{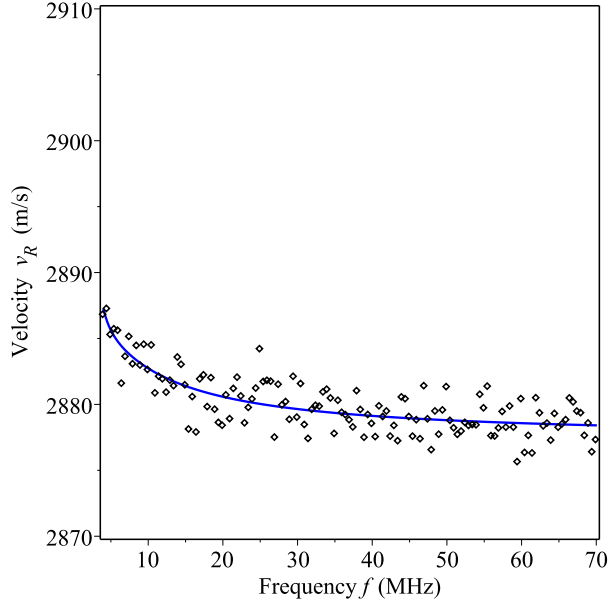}
\caption{Simulated ``experimental'' data and the  fitting curve for $\theta=90^\circ$  in the frequency window from $4$ MHz to $70$ MHz by steps of $0.5$ MHz.}
\label{simudis}
\end{center}
\end{figure}

First, by using the ``real'' relaxed stress $\Tcm(x_3)$ of the sample at state 1, we follow the method detailed in \cite{TMC} to derive theoretical dispersion curves (to third order) for the selected propagation directions $\theta$.
For instance, for  $\theta=90^\circ$ the theoretical dispersion relation is given by
  \begin{equation}
  v_R=2877.1+1.776\times10^5 \varepsilon -1.542\times10^9 \varepsilon^2+7.099\times10^{12}\varepsilon^3,
  \label{dispersionrelation}
  \end{equation}
  where $v_R$ is in m/s and $\varepsilon = 1/\textsl{k}$ is in meters. As explained in Section 5.2, in this paper we assume that the accuracy of measurement of $v_R$ is $\pm 0.1$\%. Hence for the truncated dispersion relation (\ref{asymptotic}), the approximation in replacing $v_R$ by $v_0$ in the formula $\varepsilon = v_R/(2\pi f)$ will be acceptable if $v_R-v_0$ and the correction terms $v_1/\textsl{k}$, $v_2/\textsl{k}^2$, $v_3/\textsl{k}^3$ are all within 1\% of $v_0$ (see Remark 6.2 of \cite{TMC} for further discussions). Substitution of $\varepsilon \approx v_0/(2\pi f)$ in the approximate formula (\ref{dispersionrelation}) for the phase velocity $v_R$ of the Rayleigh waves leads to the dispersion relation between $v_R$ and the frequency $f$: 
    \begin{equation}
  v_R=2877.1+\frac{2.554\times10^2}{\pi f}-\frac{3.191\times10^{3}}{\pi^2 f^2}+\frac{2.113\times10^{4}}{\pi^3f^3},
  \label{disvandf}
  \end{equation}
where $f$ is the frequency in MHz; see Figure \ref{truedis}. For each theoretical dispersion curve, we take frequency steps of $0.5$ MHz each in the frequency window from $4$ MHz to $70$ MHz and compute the theoretical values of $v_R$ at each step.   Furthermore, for each frequency $f$ in question, we assume\footnote{Based on the earlier assumption that the measurement-accuracy of $v_R$ reach $\pm0.1$\,\% and on the fact that for aluminum $v_R \approx 3,000$ m/s.} that the experimental data of $v_R$ scatter as a normal distribution with standard deviation $\sigma_{v_R}=3$ m/s about the theoretical value, and we choose a value randomly for $5$ times and then take the average as the replacement of the experimental data for $v_R$. The simulated ``experimental'' dispersion curve is obtained by using the least square method  to fit these data points with a smooth function in the form of a cubic polynomial in $1/f$. The fitting  curve and the simulated ``experimental" data for $\theta = 90^{\circ}$ are shown in Figure \ref{simudis}. The horizontal asymptotes of the fitting curves are used as the simulated ``experimental'' data for $v_0^{(1)}(\theta)$ for $\theta$ from $0^\circ$ to $180^\circ$ in steps of $15^\circ$.

As we assume that we know everything about the sample at state 0, we simply use the estimates by the formula (\ref{0orderterm}) with $\sigma_1(0) = -203.5$ MPa, $\sigma_2(0) = -412.5$ MPa, and $\zeta=10^\circ$ at the surface $x_3 =0$ as our experimental data $v_0^{(0)}$ for each $\theta$.

\subsection{Predictions and comparisons}

\begin{figure}[h!]
\begin{center}
\includegraphics[width=0.5\textwidth]{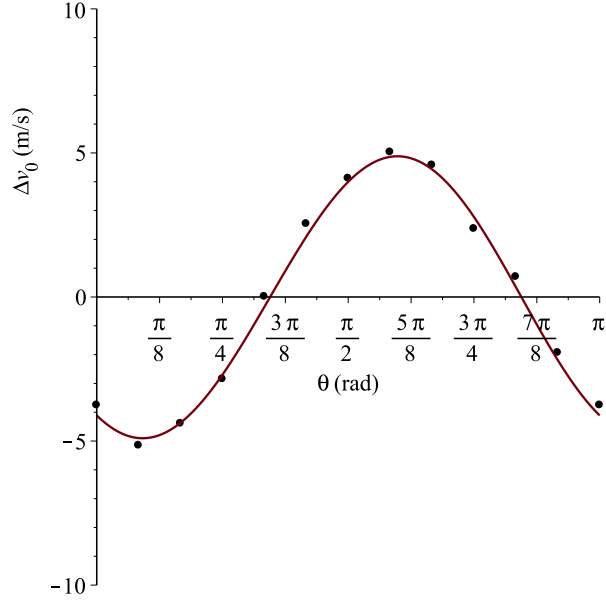}
\caption{Black dots are simulated data for  $\Delta v_0$. The fitting curve is in red.}
\label{deltavelo}
\end{center}
\end{figure}

Following the discussion in Section 4.1, we apply the Levenberg-Marquardt method to estimate the parameters $A^{(1)}, B^{(1)}, C^{(1)}$ with 1 as their initial guess to fit the simulated data on  $\Delta v_0(\theta)$ in (\ref{deltav}) for the selected $\theta$'s. The values of the parameters found are given in Table 2, and the fitting curve pertaining to these values of parameters $A^{(1)}, B^{(1)}, C^{(1)}$ are shown in Figure \ref{deltavelo}.

\begin{table}[t!]
\begin{center}
\begin{tabular}{|c|c|c|c|c|c|c|c|c|}
  \hline
   $A^{(1)}$ & $B^{(1)}$ & $C^{(1)}$ \\
  \hline
   $-2.32763\,10^2$ & $-4.16791\,10^1$ & $-7.88577\,10^1$ \\
   \hline
\end{tabular}
\end{center}
\caption{Fitting values of parameters $A^{(1)}, B^{(1)}, C^{(1)}$ in units of MPa as determined from simulated data on $\Delta v_0$.}  \label{Table2}
\end{table}

Substituting the fitted values of Table \ref{Table2} into (\ref{unkowns}), we obtain $\Tcc_m(0)=-232.8$ MPa, $\Tcc_d(0)=-89.2$ MPa and 
$\zeta=31.1^\circ$ at the surface $x_3 = 0$ for the sample at state 1. Consequently, we have $\sigma_1(0)=-143.6$ MPa and 
$\sigma_2(0)=-322.0$ MPa at the free surface. Then the surface prestress $\Tcm(0)$ can be derived from (\ref{rotation}).  Thus we have $\Tcc_{11}(0)=-274.4$ MPa, $\Tcc_{22}(0)=-191.1$ MPa and $\Tcc_{12}(0)=-78.9$ MPa.  Substituting the surface stress $\Tcm(0)$ into (\ref{cubicfit}) and applying the algorithm given in \cite{TMC}, we get $v_1, v_2, v_3$ in terms of the parameters $a_m, b_m, c_m$ 
$(m=1,2,3)$.  Following the discussion in Section 4.2, we repeat the calculations for the cases $\theta =0^\circ \text{ and } 45^\circ$. The process is the same for different $\theta$, except that we should use the formulas given in Appendix A for the components (with respect to the spatial coordinate system) of $\Phi$, $\Psi^{(i)}(i=1,\dots,4)$ and $\Theta$ in (\ref{constitutive}) for the specific $\theta$ in question .

To get better estimates of $a_m, b_m, c_m$ ($m=1,2,3$), we consider 3 groups of $\theta$: (1) $\theta=0^\circ, 45^\circ, 90^\circ$; (2) $\theta=60^\circ, 120^\circ, 180^\circ$; (3) $\theta=30^\circ, 105^\circ, 150^\circ$. Group 1 has been discussed above. The other two groups are processed in the same way.
Table \ref{tablebi}  shows the results of $a_m, b_m, c_m$ $(i=1,2,3)$ for these three groups and the corresponding average values.
\begin{table}[h!]
\begin{center}
\begin{tabular}{|c|c|c|c|c|c|c|c|c|c|}

  \hline
  & $a_1$ &$a_2$&$a_3$  \\
  \hline
    group (1) &$9.932275194\, 10^{2}$ & $2.050029714\, 10^{3}$ & $5.699045777\,10^{2}$ \\
   \hline
   group (2)&  $9.958481153\, 10^{2}$ & $2.052999597\, 10^{3}$ & $5.650730529\,10^{2}$ \\
   \hline
   group (3)&  $9.939524939\, 10^{2}$ & $2.046795586\, 10^{3}$ & $5.417913915\,10^{2}$\\
   \hline
   average  &  $9.943427095\, 10^{2}$ & $2.049941632\, 10^{3}$ & $5.589230074\,10^{2}$\\
    \hline
  & $b_1$ & $b_2$ & $b_3$  \\
  \hline
   group (1)&$1.414984597\, 10^{3}$ & $2.69694662\, 10^{3}$ & $9.564504997\,10^{2}$ \\
   \hline
   group (2)&$1.418226166\, 10^{3}$ & $2.70135238\, 10^{3}$ & $9.496635036\,10^{2}$ \\
   \hline
   group (3)&$1.416637377\, 10^{3}$ & $2.70149117\, 10^{3}$ & $9.601827370\,10^{2}$\\
   \hline
   average  &$1.416616047\, 10^{3}$ & $2.69993006\, 10^{3}$ &  $9.554322468\,10^{2}$\\
  \hline

  &$c_1$&$c_2$&$c_3$  \\
  \hline
   group (1)  &$-2.213326631\, 10^{2}$ & $-2.136493828\, 10^{2}$ & $-6.64566032\,10^{1}$ \\
   \hline
   group (2)  &$-2.203477994\, 10^{2}$ & $-2.127819162\, 10^{2}$ & $-7.34684457\,10^{1}$ \\
   \hline
   group (3)  &$-2.203303837\, 10^{2}$ & $-2.131984256\, 10^{2}$ & $-7.62330586\,10^{1}$\\
   \hline
   average    &$-2.206702821\, 10^{2}$ & $-2.132099082\, 10^{2}$ & $-7.20527025\,10^{1}$\\
   \hline
\end{tabular}
\end{center}
\caption{Values and average of $a_1, b_1, c_1$  (MPa/mm), $a_2, b_2, c_2$ (MPa/mm$^2)$, $a_3, b_3, c_3$ (MPa/mm$^3)$ from three groups of $\theta$: (1) $\theta=0^\circ, 45^\circ, 90^\circ$; (2) $\theta=60^\circ, 120^\circ, 180^\circ$; (3) $\theta=30^\circ, 105^\circ, 150^\circ$.}
\label{tablebi}
\end{table}

\begin{figure}[t!]
\centering
\subfigure[][$\Tcff_{11}$]{\epsfig{figure=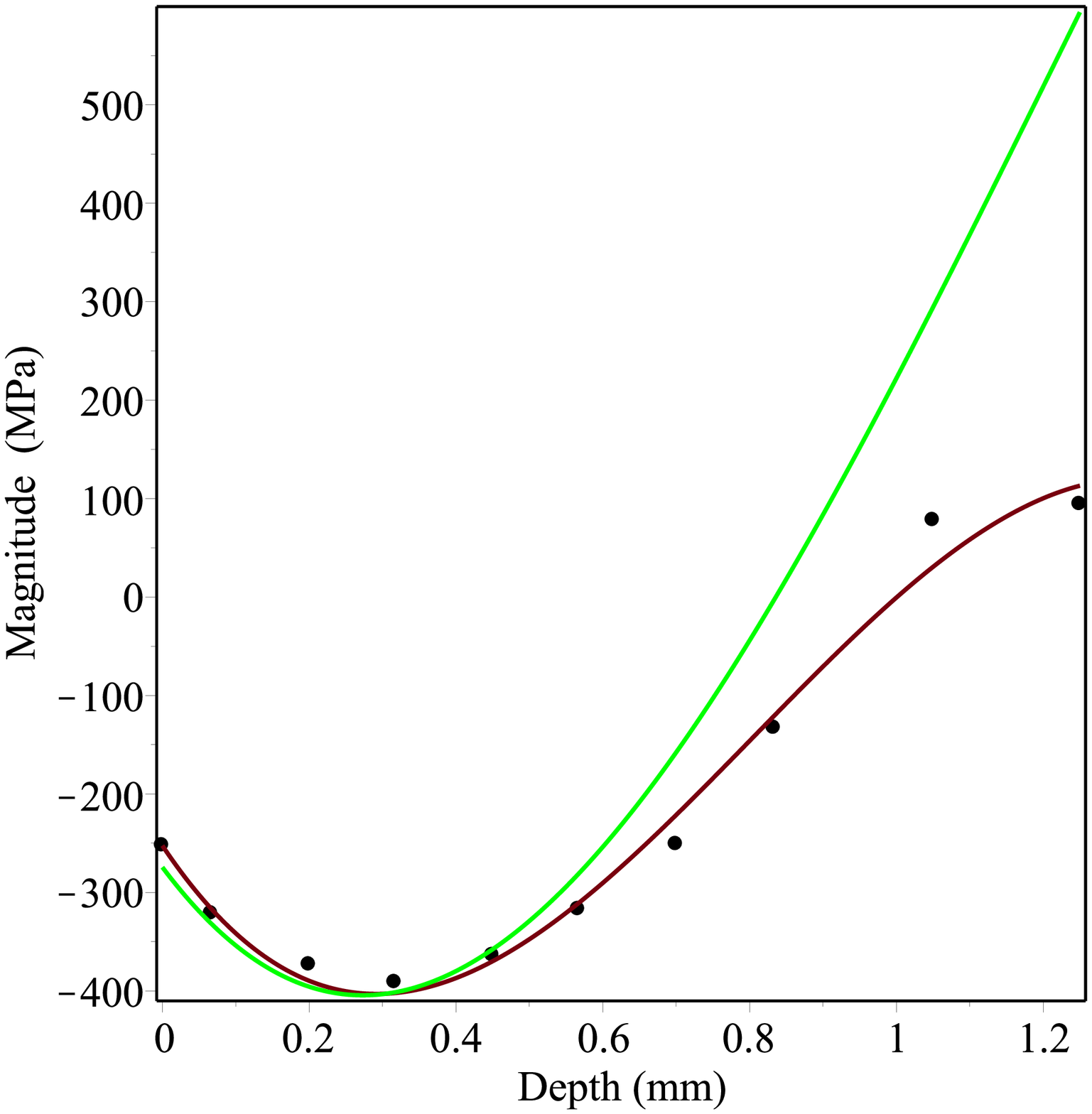 , width=1.9in}}~
\subfigure[][$\Tcff_{22}$]{\epsfig{figure=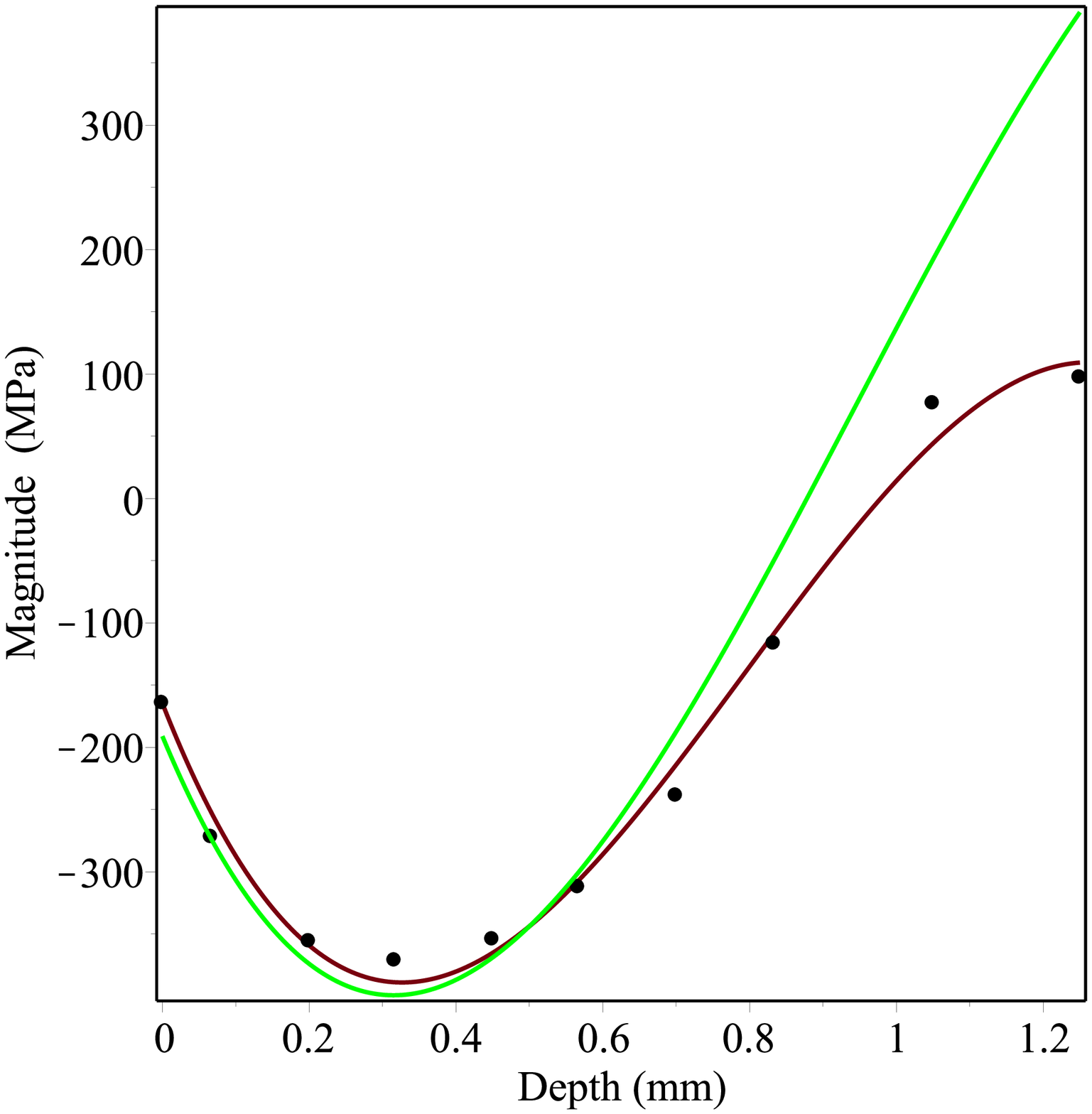 , width=1.9in}}~
\subfigure[][$\Tcff_{12}$]{\epsfig{figure=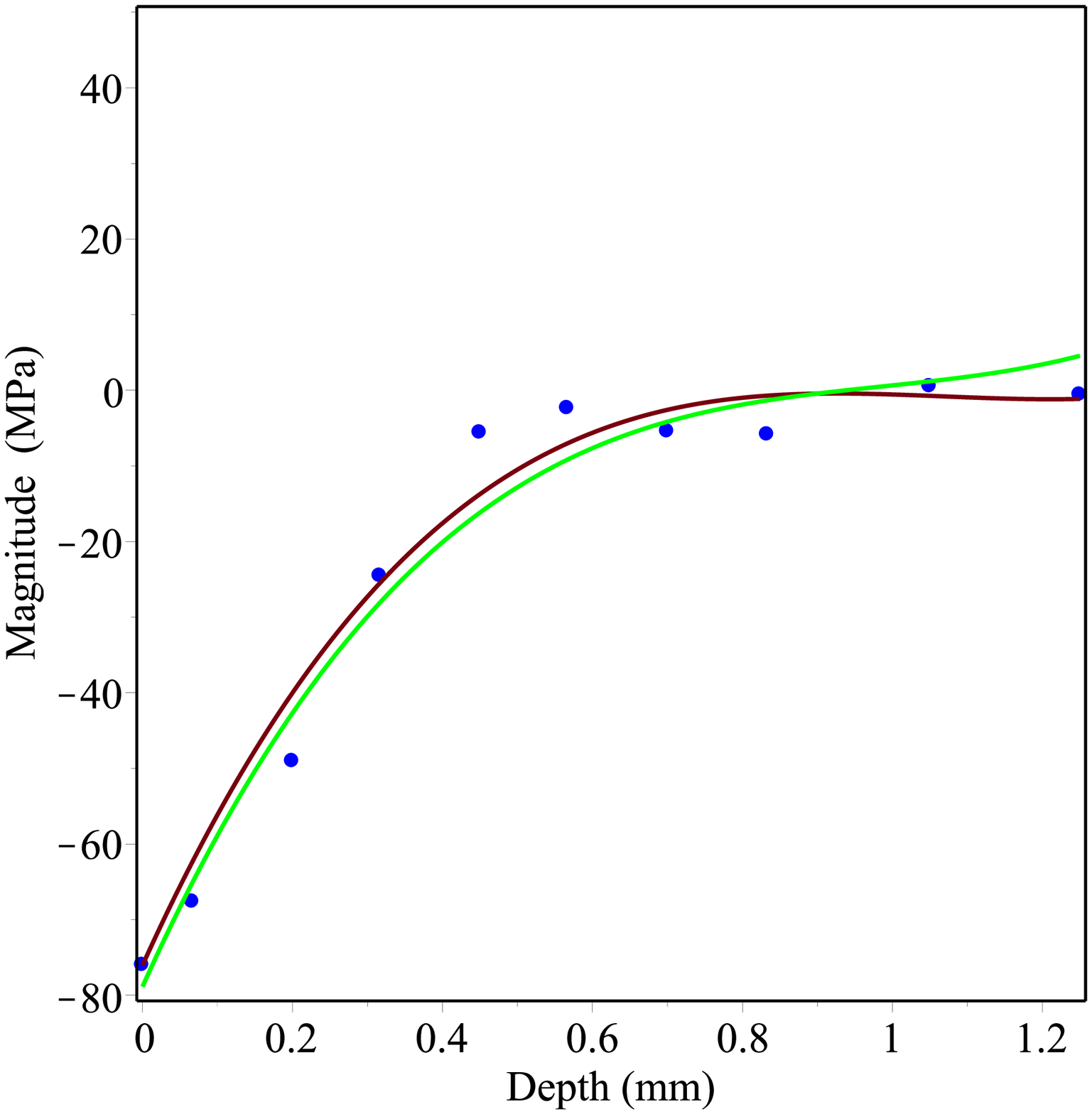 , width=1.9in}}
\caption{Comparison between the components $\protect\Tcf_{ij}$ ($ij\in\{11,22,12\}$) of the simulated prestress and of the ``real'' prestress in in the sample at state 1.}
\label{stresscomp1}
\end{figure}

\begin{figure}[!htb]
\centering
\subfigure[][Principal stress $\sigma_1$]{\epsfig{figure=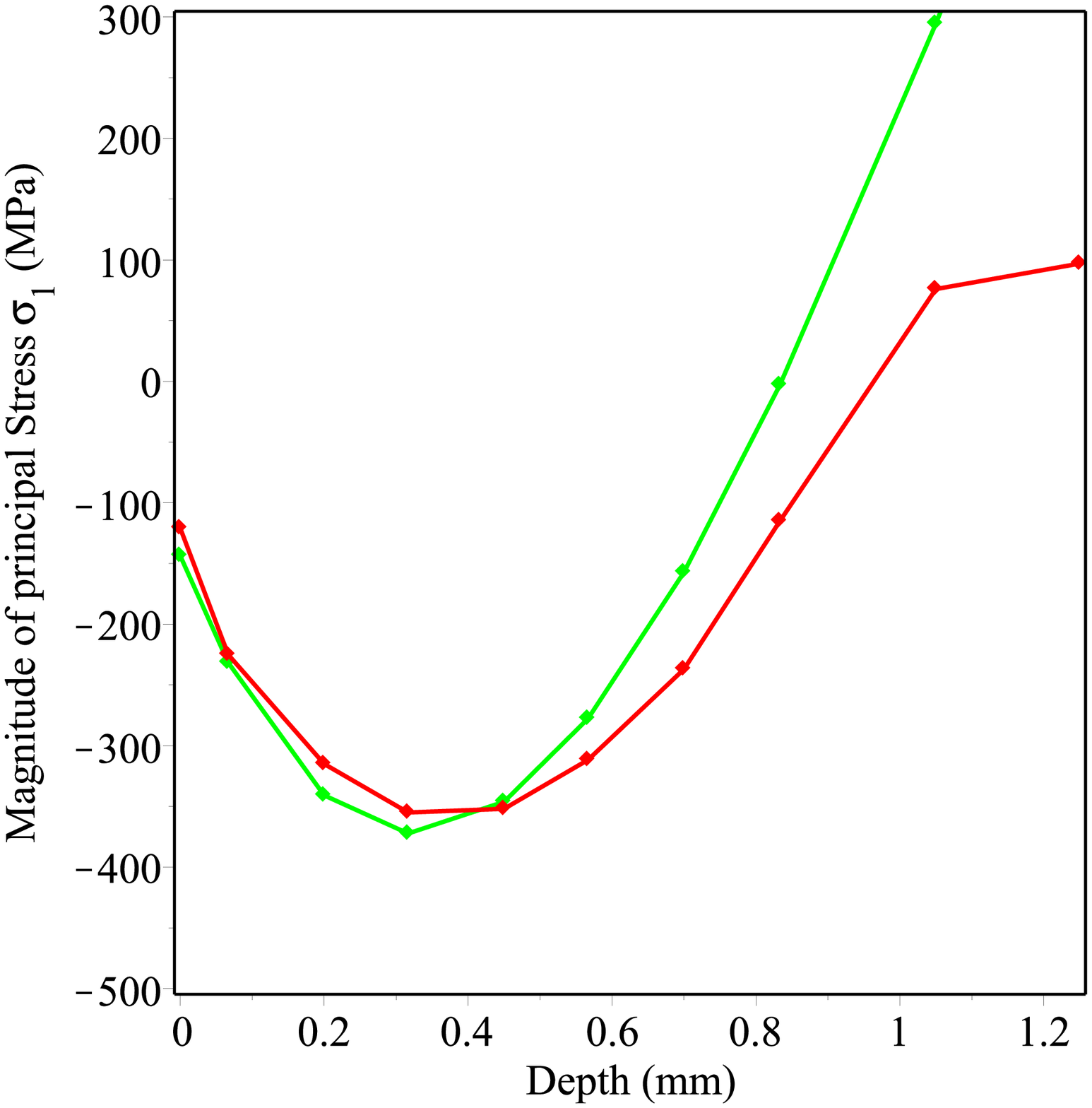 , width=2.1in}}~
\subfigure[][Principal stress $\sigma_2$]{\epsfig{figure=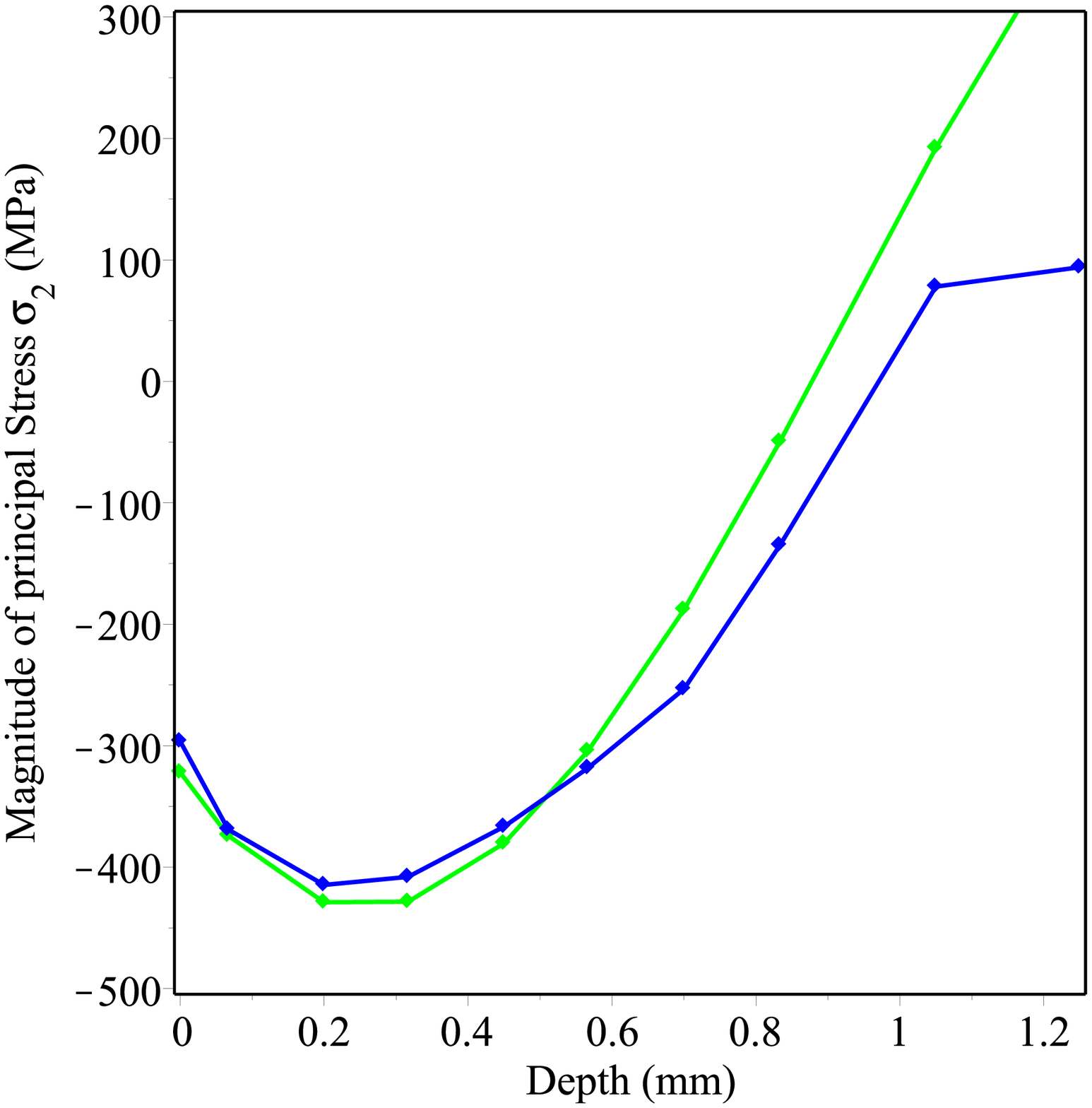 , width=2.1in}}
\caption{Comparison between the simulated principal prestresses and the ``real" principal prestresses in state 1. The green curves are the simulated principal prestresses while the red one the ``real" principal stress $\sigma_1$ and the blue one $\sigma_2$.
}
\label{principal}
\end{figure}

We use the average values of $a_m, b_m, c_m$ as our simulated results.  The components of the corresponding  $\Tcm$ are shown below:
\begin{align}
\Tcc_{11} &=-274.4+9.943\times10^{2}x_3+2.050\times10^{3}x_3^2+5.589\times10^{2}x_3^3,\nonumber\\
\Tcc_{22}&=-191.1+1.417\times10^{3}x_3+2.700\times10^{3}x_3^2+9.554\times10^{2}x_3^3,\label{fittingcurve}\\
\Tcc_{12}&=-78.9-2.207\times10^{2}x_3-2.132\times10^{2}x_3^2-7.205\times10^{1}x_3^3,\nonumber
\end{align}
where the stresses are in MPa and $-x_3 \geq 0$ denotes the depth in units of mm.
Comparisons between the components of our simulated $\Tcm(x_3)$ (green curves) and those of the ``real'' residual stress (black dots) in the sample at state 1 are shown in Figure \ref{stresscomp1}, where the red curves are the fitting curves for the ``real'' stresses.
 A comparison of the corresponding principal stresses is illustrated in Figure \ref{principal}.

\section{The second-order approximation}

\begin{figure}[h!]
\begin{center}
\includegraphics[width=0.5\textwidth]{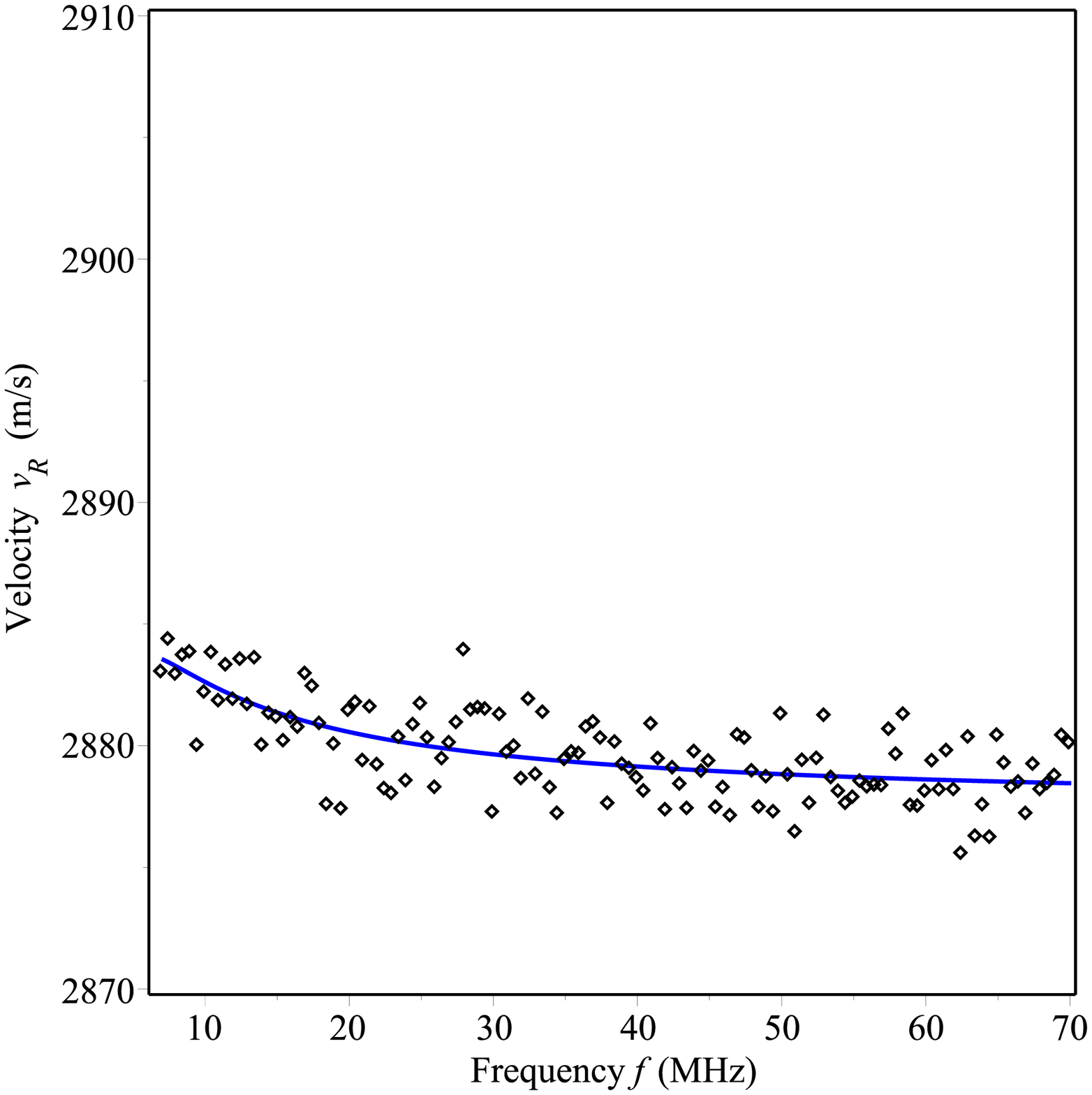}
\caption{Simulated data and the $2$nd-order fitting curve for $\theta=90^\circ$  in the frequency window from $7$ MHz to $70$ MHz by steps of $0.5$ MHz.}
\label{simudisQ}
\end{center}
\end{figure}

In some applications, information on the stress from the surface to a depth of about $0.5$ mm would be sufficient. Moreover, in Section \ref{testsample} the lower bound of the frequency window is $4$ MHz, which could be too low 
in practice because diffraction errors are much larger for frequencies lower than $4$ or $5$ MHz.  In this section we will truncate the asymptotic expansion (\ref{nth-v-esp}) for the Rayleigh-wave velocity $v_R$ at the order $\varepsilon^2$ and use a frequency window of lower boundary $7$ MHz.
The same aluminum sample as  in Section \ref{testsample} is considered.   
The depth profiles of residual stresses $\Tcmo$\hspace*{0.4cm}$(x_3)$ and $\Tcm(x_3)$ for state 0 and 1, respectively, are the same as Section \ref{testsample}. However, here we focus only on the parts corresponding to the range of $-x_3$ in $[0, 0.6]$ mm.  
The simulated ``experimental" data on $v_R$ at various frequencies for various propagation directions $\theta$ are the same as in Section \ref{testsample}, but we will only use the data within the frequency window from $7$ MHz to $70$ MHz.
Unlike the third-order approximation in Section \ref{testsample}, here we obtain the dispersion curves  by applying the least square method to fit the simulated data points with the quadratic form
\begin{eqnarray}
v_R(\theta) = v_0(\theta) +v_1(\theta)\varepsilon+v_2(\theta){\varepsilon}^2, \qquad \mbox{where $\varepsilon = 1/\textsl{k}$.}
\label{quadraticapp}
\end{eqnarray}
As illustration, the fitting curve and the simulated ``experimental" data for $\theta=90^\circ$ are shown in Figure \ref{simudisQ}. 
The horizontal asymptotes of the fitting curves are treated as the simulated data for $v_0^{(1)}$ for $\theta$ from $0^\circ$ to $180^\circ$ in steps of $15^\circ$.

Just as what we did in Section \ref{testsample}, the fitted values of the parameters $A^{(1)}, B^{(1)}, C^{(1)}$ in (\ref{deltav}) are determined in the same way. Here, in the second-order approximation,  we have $A^{(1)}=-231.1$ MPa, $B^{(1)}=-41.1$ MPa, $C^{(1)}=-73.9$ MPa. 
Therefore from (\ref{unkowns}), we obtain $\Tcc_m(0)=-231.1$ MPa, $\Tcc_d(0)=-84.6$ MPa and $\zeta=30.5^\circ$ at the surface $x_3=0$ of the sample at state 1. It follows that $\sigma_1(0)=-146.6$ MPa and $\sigma_2(0)=-315.7$ MPa at the free surface. The surface residual stress $\Tcm(0)$ has the components $\Tcc_{11}(0)=-272.2$ MPa, $\Tcc_{22}(0)=-190.1$ MPa and $\Tcc_{12}(0)=-73.9$ MPa. 

\begin{figure}[t!]
\centering
\subfigure[][$\Tcff_{11}$]{\epsfig{figure=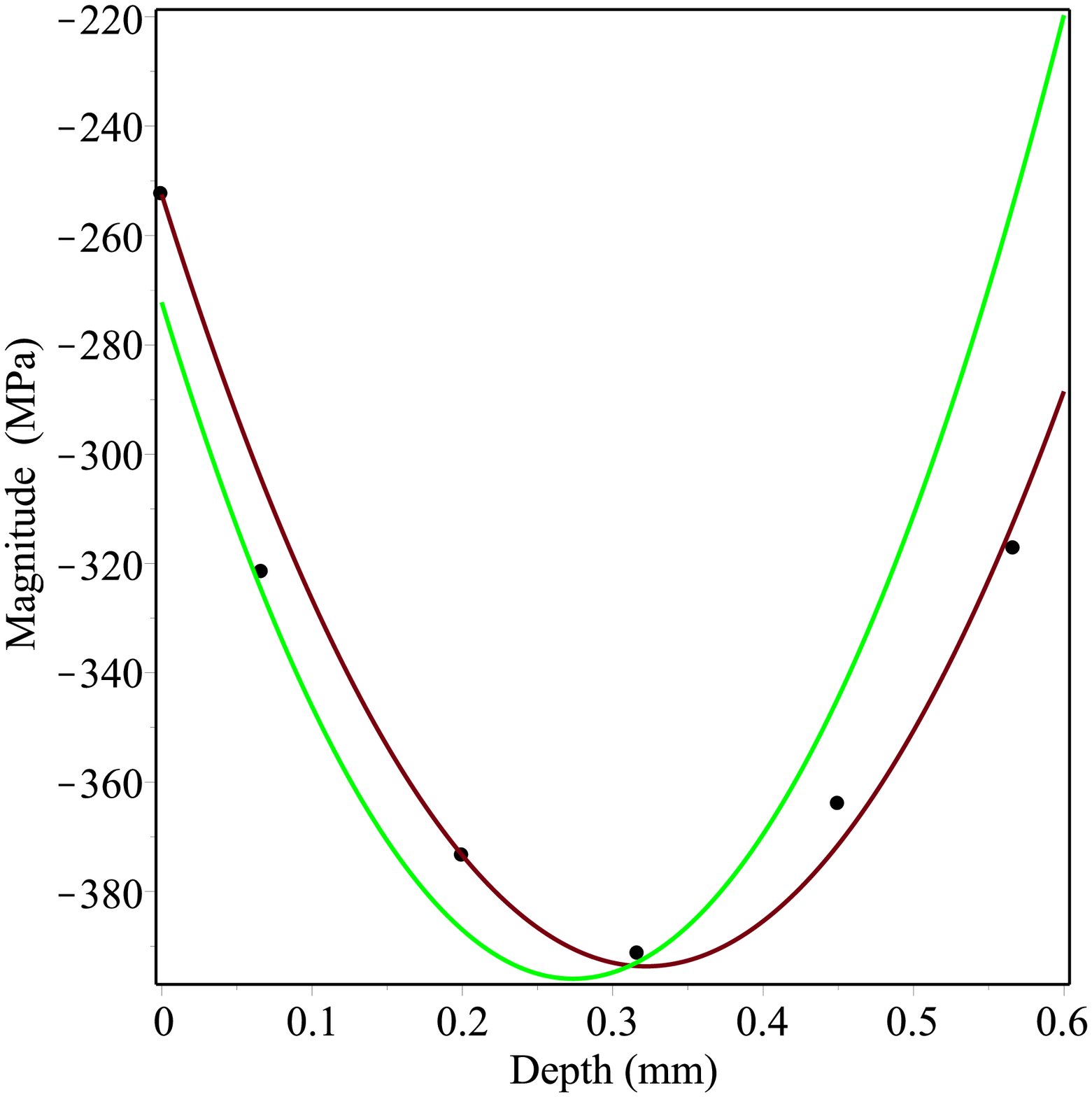 , width=1.9in}}~
\subfigure[][$\Tcff_{22}$]{\epsfig{figure=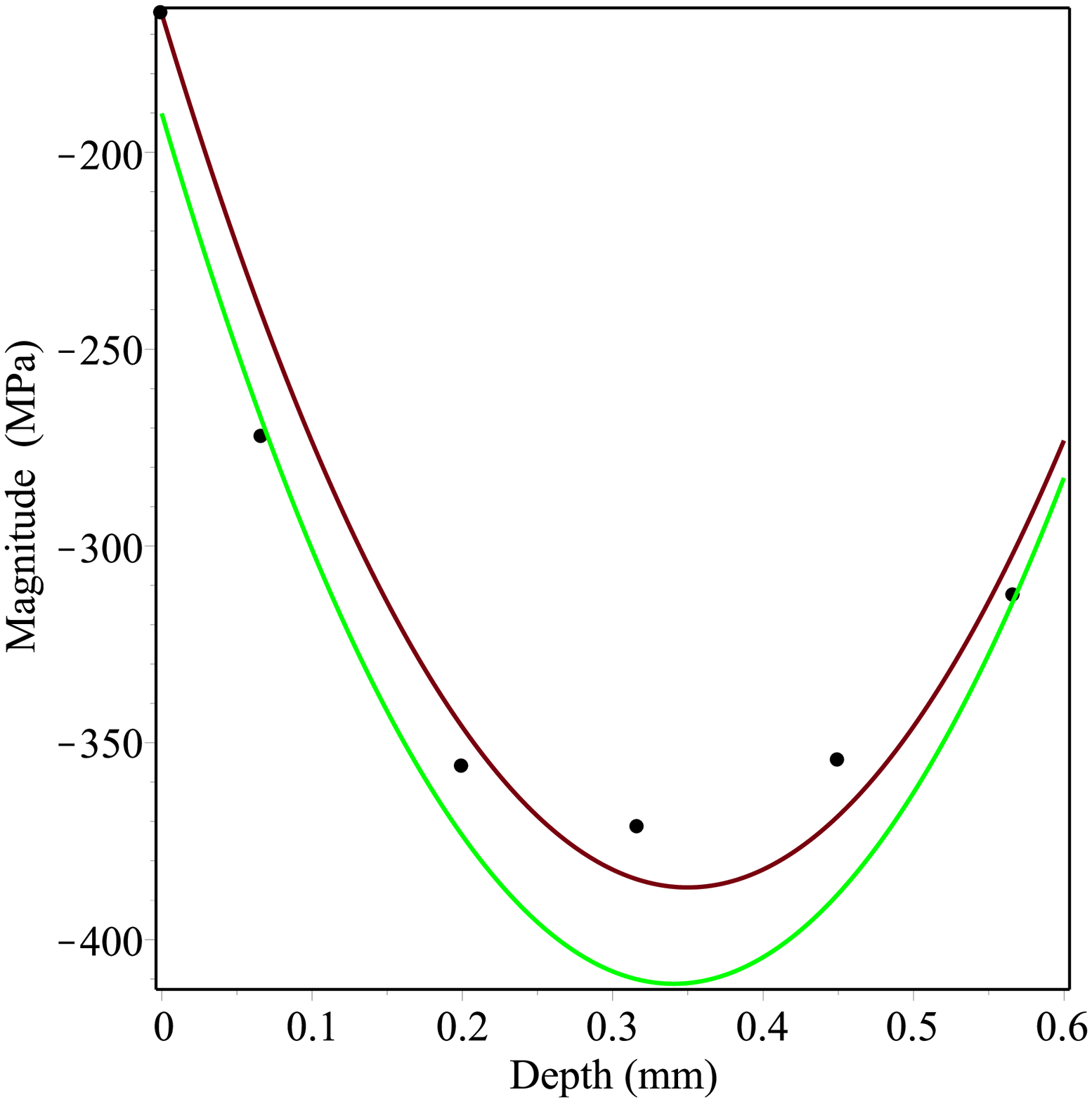 , width=1.9in}}~
\subfigure[][$\Tcff_{12}$]{\epsfig{figure=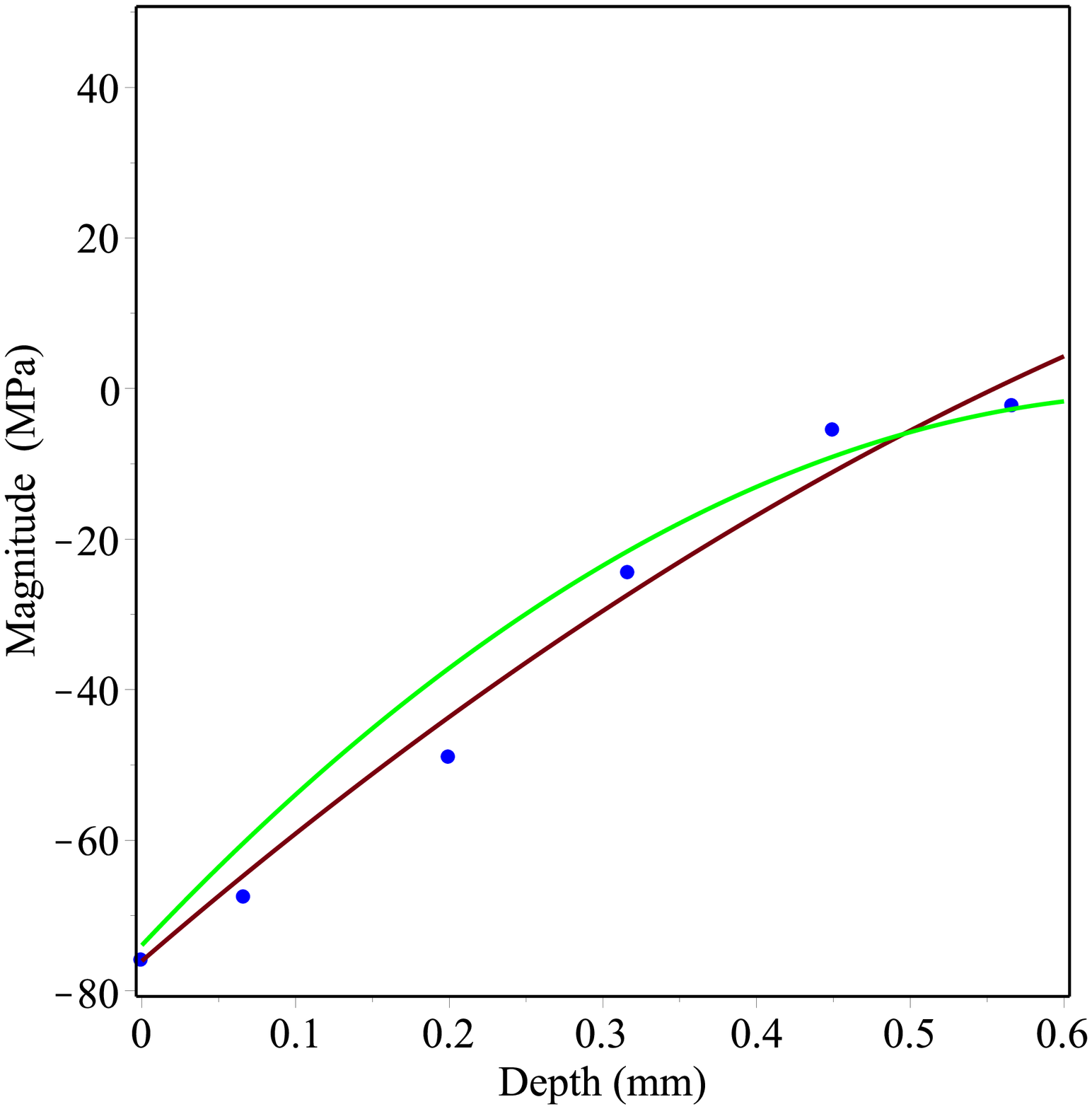 , width=1.9in}}\\
\caption{Comparison between the simulated $\protect\Tcf_{ij}$ where $ij\in\{11,22,12\}$ and the original prestress component in state 1 for the frequency window from $7$ MHz to $70$ MHz.
}
\label{stresscompnorelaxQ}
\end{figure}

\begin{figure}[!htb]
\centering
\subfigure[][$\sigma_1$]{\epsfig{figure=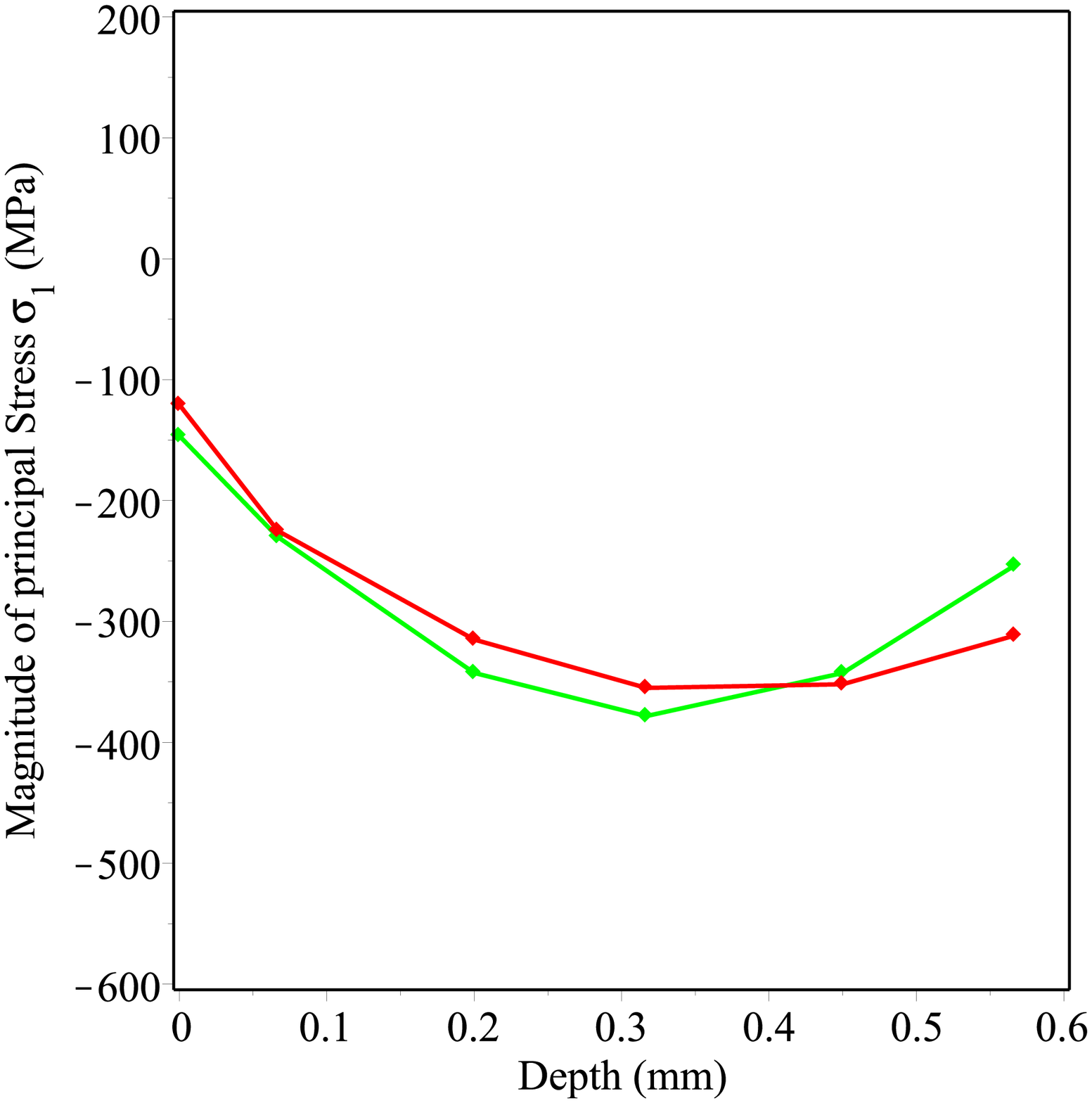 , width=1.9in}}~
\subfigure[][$\sigma_2$]{\epsfig{figure=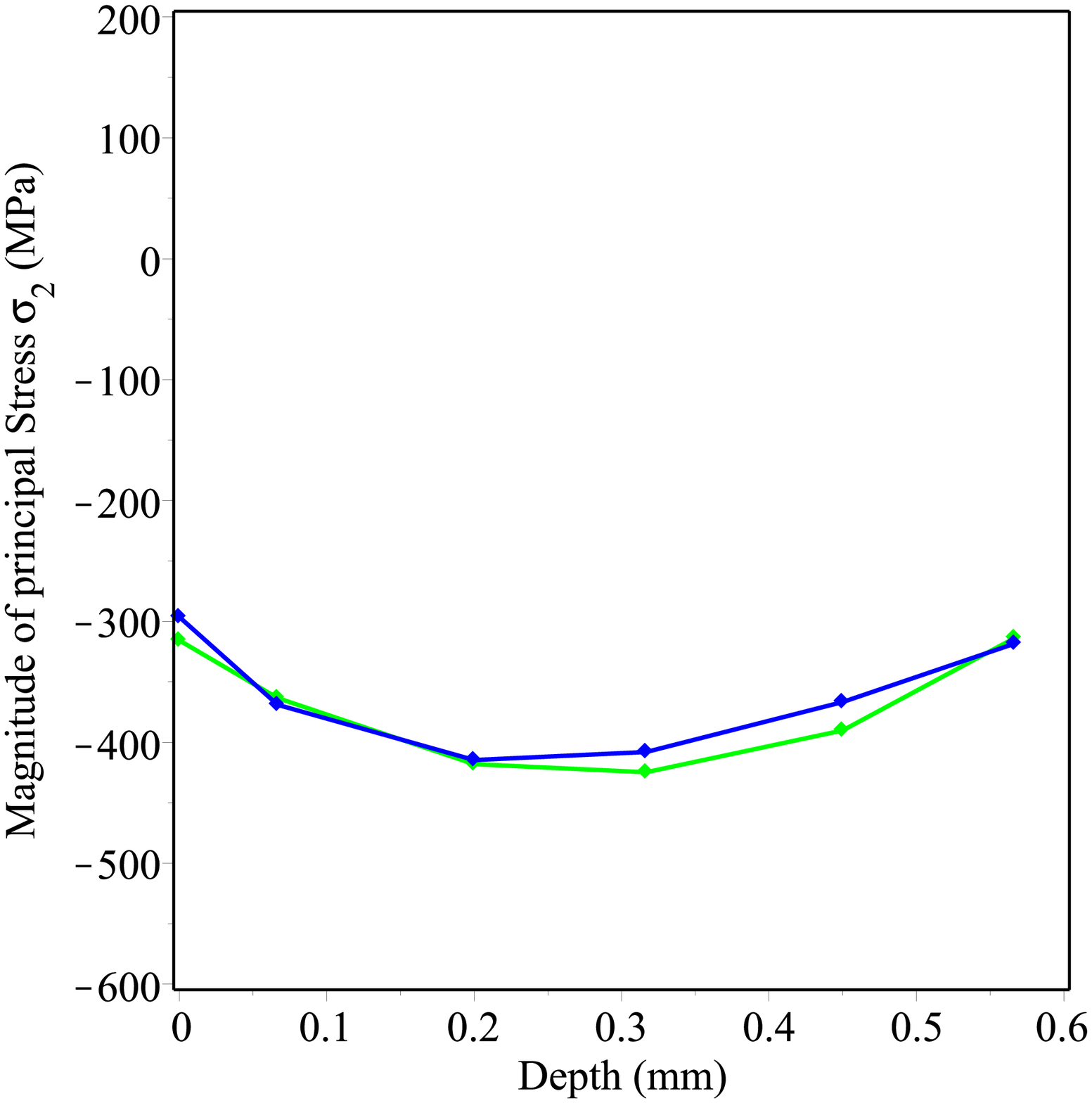 , width=1.9in}}
\caption{Comparison between the simulated principal prestress (the green curves) and the ``real" principal prestress  in state 1 for the frequency window from $7$ MHz to $70$ MHz}
\label{principalstressrelaxQ}
\end{figure}

Here we assume that the components of the relaxed prestress can be fitted by some quadratic forms
\begin{align}
\Tcc_{11}&=\Tcc_{11}(0)+\hat{a}_1x_3+\hat{a}_2x_3^2,\nonumber\\
\Tcc_{22}&=\Tcc_{22}(0)+\hat{b}_1x_3+\hat{b}_2x_3^2,\label{quadraticfit}\\
\Tcc_{12}&=\Tcc_{12}(0)+\hat{c}_1x_3+\hat{c}_2x_3^2.\nonumber
\end{align}
For the second-order approximation, we just need to follow the algorithm given in \cite{TMC} to determine $v_1, v_2$ in terms of $\hat{a}_m, \hat{b}_m,\hat{c}_m$ $(m=1,2)$. From Corollary 3.2, $v_1$ is of first-order in $\hat{a}_1,\hat{b}_1,\hat{c}_1$ and $v_2$ is of first-order in $\hat{a}_2,\hat{b}_2,\hat{c}_2$.
Just as what we did in Section \ref{testsample}, we use waves of three different propagation directions, say $\theta=0^\circ, 45^\circ, 90^\circ$, and compare the quadratic-fitting dispersion curves in the form of (\ref{quadraticapp}) with the parametric dispersion curves $v_R=v_0+v_1(\hat{a}_1,\hat{b}_1,\hat{c}_1)\,\varepsilon+v_2(\hat{a}_1,\hat{b}_1,\hat{c}_1, \hat{a}_2,\hat{b}_2, {c}_2)\,\varepsilon^2$ to determine $\hat{a}_1,\hat{b}_1,\hat{c}_1$ first, and then $\hat{a}_2,\hat{b}_2,\hat{c}_2$. For instance, the components of the corresponding  $\Tcm(x_3)$ as obtained from data that pertain to the propagation directions $\theta=0^\circ, 45^\circ, 90^\circ$ are:
\begin{align}
\Tcc_{11} &=-272.2+9.048\times10^{2}x_3+1.654\times10^{3}x_3^2,\nonumber\\
\Tcc_{22}&=-190.1+1.299\times10^{3}x_3+1.908\times10^{3}x_3^2,\label{fittingcurveQ}\\
\Tcc_{12}&=-73.9-2.159\times10^{2}x_3-1.591\times10^{2}x_3^2,\nonumber
\end{align}
where $\Tcc_{ij}(x_3)$ are in units of MPa and $-x_3 \geq 0$ denotes the depth in units of mm.
A comparison between our simulated prestresses (green curves) and the ``real'' prestresses (black dots) in state 1 are shown in Figure \ref{stresscompnorelaxQ}, where the red curves are the fitting curves for the ``real'' prestresses. 
A comparison of the corresponding principal stresses is illustrated in Figure \ref{principalstressrelaxQ}, where the green curves are the simulated principal prestresses. From Figure \ref{stresscompnorelaxQ} and \ref{principalstressrelaxQ}, we see that the quadratic approximation gives good estimates of the stress profiles up to a depth of about $0.5$ mm.

\section{Closing remarks}

In this paper we study the inverse problem on inferring depth profile of near-surface residual stress in a weakly anisotropic medium by using the algorithm given in \cite{TMC} for finding each term of a high-frequency asymptotic formula for Rayleigh-wave dispersion. We show that, after the zeroth order terms  are determined, Theorem 3.1 and Corollary 3.2 reduce the inverse problem to the routine work of solving iteratively systems of linear equations. We apply the theory to the practical problem on monitoring retention of residual stress induced by the surface-conditioning treatment of low plasticity burnishing on an aluminum thick plate sample if all other material parameters (including texture coefficients) remain unchanged. Our study suggests that, if measurement of Rayleigh-wave velocity has an accuracy of $\pm 0.1$\%, then the depth profile of the residual stress from the surface to a depth of 0.5 to 0.7 mm---with the upper limit depending on the frequency window where velocity measurements have $\pm 0.1$\% accuracy---can be recovered.

In practice, besides accuracy of the velocity measurements, there is one more crucial question to be answered for our proposed method to be applicable. It is whether we can identify a frequency window $[f_m, f_M]$ which satisfies the following three conditions: (i) $f_m$ should be high enough for some computable truncated version of the high-frequency formula derived in \cite{TMC} to be valid; (ii) $f_M$  should be sufficiently low that the effects of surface roughness on Rayleigh-wave dispersion can be ignored for waves with frequencies within the window; (iii) dispersion of Rayleigh waves with frequencies within the window should be sufficiently pronounced that information on near-surface stress can be extracted from the dispersion data. All these depend on the specific material medium and sample in question. For example, condition (iii) depends most notably on the size of the acoustoelastic effect for Rayleigh waves propagating in that medium, and condition (ii) depends on the surface finish of the sample. Hence whether the proposed method would work for a specific application can only be decided after careful study on a case-by-case basis.

\section*{Acknowledgements}
Chen conducted his work in this research project while he was a postdoctoral scholar at Department of Mathematics, University of Kentucky, Lexington, USA. The research efforts of Tanuma were partially supported by JSPS KAKENHI Grant Number JP26400157.
The research of Kube reported in this paper was performed under contract(s)/instrument(s) W911QX-16-D-0014 with the U.S. Army Research Laboratory. The views and conclusions contained in this paper are those of the authors, Bennett Aerospace and the U.S. Army Research Laboratory. Citation of manufacturer's or trade names does not constitute an official endorsement or approval of the use thereof. The U.S. Government is authorized to reproduce and distribute reprints for Government purposes notwithstanding any copyright notation hereon.

\appendix
\section{Details on constitutive equation of 7075-T651 aluminum sample}
 \label{7075-T651}
 
 In this appendix we provide the details that complete the constitutive equation of the 7075-T651 aluminum sample studied in this paper.
 
\subsection{Material parameters}
 
In our computations we take $\lambda = 60.79$ GPa and $\mu=26.9$ GPa,  which correspond to the mean values of $\mu$ and Young's modulus $E = 71.43$ GPa obtained by Radovic et al.\ \cite{RLR} in their RUS (resonant ultrasound spectroscopy) measurements on sixteen 7075-T651 samples. As for the other 10 parameters, we are not aware of any experimentally determined value reported in the literature. Hence we adopt the values predicted by  the Man-Paroni model (\cite{Man,MP,PM}) from second-order and third-order elastic constants of single-crystal pure aluminum reported by  \cite{Thomas} and  \cite{SR}, respectively: $\alpha=-16.49$ GPa, $\beta_1=0.89$, $\beta_2=0.96$, $\beta_3=-2.63$, $\beta_4=-4.54$, $\tilde{b}_1=-3.32$, $\tilde{b}_2=-0.61$, $\tilde{b}_3=0.14$, $\tilde{b}_4=1.54$ and $a=12.10$.

\subsection{Texture coefficients}
 
 The texture coefficients of the sample that pertain to the treated surface and several depths (up to 0.225 mm) were determined by X-ray diffraction and serial sectioning. They were found to be largely constant for the planes examined. In this paper we simply take the texture coefficients to be constant for the entire sample. The values which refer to the material $OX'Y'Z'$ coordinate system are:
 \begin{itemize}
 \item $W'_{400}=0.00393$, $W'_{420}=-0.00083$, $W'_{440}=-0.00233$, $W'_{600}=0.00025$, $W'_{620}=-0.0004$, $W'_{640}=-0.00033$, and $W'_{660}=0.00035$.
\end{itemize}

 \subsection{Components of tensors $\bphi$, $\bth$, and $\bpsi$}
 
 All components of tensors below refer to the coordinate system $OXYZ$ defined in Section 2. The material coordinate system $OX'Y'Z'$ has its $OZ'$-axis agree with the $OZ$-axis. Let $\theta$ be the angle of rotation about the $OZ$ axis which brings the $OX$ axis to the $OX'$ axis (see Fig.\ 1 in Section 4).

An $r$-th order tensor $\Hm$\ is said to be harmonic if it is totally symmetric
and traceless, i.e., its components
$H_{i_1 i_2 \cdots i_r}$ satisfy
$H_{i_1 i_2 \cdots i_r} = H_{i_{\tau(1)} i_{\tau(2)} \cdots i_{\tau(r)}}$ 
for each permutation $\tau$ of $\{1, 2, ..., r\}$, and
$  \text{tr}_{j,k}\Hm = \zero $
for any pair of distinct indices $j$ and $k$. For example, for $r=3$ we have $H_{112} = H_{121} = H_{211}$, etc.\ from total symmetry,
and $H_{111} + H_{212} + H_{313} = 0$, etc.\ from the traceless condition.

The fourth-order tensor $\bphi$ and the sixth-order tensor $\bth$
are harmonic. All the non-trivial components of $\bphi$ can be obtained from the following five through
the total symmetry of and the traceless condition on the harmonic tensor $\bphi$: 
\begin{gather*}
\Phi_{1122}  =  W'_{400} - \sqrt{70}\,W'_{440}\cos4\theta, \quad
\Phi_{1133}  = -4W'_{400} + 2\sqrt{10}\,W'_{420}\cos2\theta, \\
\Phi_{2233}  = -4W'_{400} - 2\sqrt{10}\,W'_{420}\cos2\theta,  \quad
\Phi_{1112}  = -\sqrt{10}\,W'_{420}\sin2\theta + \sqrt{70}\,W'_{440}\sin4\theta, \\
\Phi_{2212} = -\sqrt{10}\,W'_{420}\sin2\theta - \sqrt{70}\,W'_{440}\sin4\theta.   
\end{gather*}

The non-trivial components of $\bth$\ can be obtained from the
following seven by using the total symmetry of and the traceless condition
on $\bth$:
 \begin{xxalignat}{2}
\hspace*{0.4in}\Theta_{222211} & = -W'_{600} - \frac{\sqrt{105}}{15}\,W'_{620}\cos2\theta
  + \sqrt{14}\,W'_{640}\cos4\theta + \sqrt{231}\,W'_{660}\cos6\theta, &  \\  
\Theta_{222233} & = 6W'_{600} + \frac{16\sqrt{105}}{15}\,W'_{620}\cos2\theta
  + 2\sqrt{14}\,W'_{640}\cos4\theta, & \\  
\Theta_{333311} & = -8W'_{600} + \frac{16\sqrt{105}}{15}\,W'_{620}\cos2\theta, \qquad
\Theta_{333322}  = -8W'_{600} - \frac{16\sqrt{105}}{15}\,W'_{620}\cos2\theta, &  \\   
\Theta_{122222} & = \frac{\sqrt{105}}{3}\,W'_{620}\sin2\theta
+ 2\sqrt{14}\,W'_{640}\sin4\theta + \sqrt{231}\,W'_{660}\sin6\theta, &  \\ 
\Theta_{122233} & =  -\frac{8\sqrt{105}}{15}\,W'_{620}\sin2\theta
- 2\sqrt{14}\,W'_{640}\sin4\theta,  \qquad
\Theta_{123333}  = \frac{16\sqrt{105}}{15}\,W'_{620}\sin2\theta. &   
\end{xxalignat}

The components of the sixth-order tensors $\bpsi^{(i)}(w)$ 
are given in terms of those of the harmonic
tensor $\bphi$ by the following
formulae:
\begin{align*}
\Psi^{(1)}_{ijklmn}  & =  \Phi_{ijkl}\delta_{mn}, \qquad
\Psi^{(2)}_{ijklmn}  =  \Phi_{klmn}\delta_{ij} + \Phi_{ijmn}\delta_{kl}, \\
\Psi^{(3)}_{ijklmn} & =  \delta_{ik}\Phi_{jlmn} + \delta_{il}\Phi_{jkmn}
 + \delta_{jk}\Phi_{ilmn} + \delta_{jl}\Phi_{ikmn},  \\  
\Psi^{(4)}_{ijklmn} & =  \delta_{im}\Phi_{jnkl} + \delta_{in}\Phi_{jmkl}
 + \delta_{jm}\Phi_{inkl} + \delta_{jn}\Phi_{imkl} 
    + \delta_{km}\Phi_{lnij} + \delta_{kn}\Phi_{lmij}
 + \delta_{lm}\Phi_{knij} + \delta_{ln}\Phi_{kmij}, 
\end{align*}
where $\delta_{ij}$ is the Kronecker delta.

\section{Proof of Theorem 3.1} 

To prove Theorem 3.1,  we start by observing how  ${\bf Z}_m\  (m=1,2,\cdots, n)$ are affected by the first and the higher order  $x_3$-derivatives of $\Tcm(x_3)$ at $x_3=0$.

\noindent
\begin{lem}
For $m=1,2,\cdots, n$,
${\bf Z}_m$ depends on $\Tcm(0)$ and on the  $x_3$-derivatives of $\Tcm(x_3)$ at $x_3=0$ up to those of order $m$; in particular,  ${\bf Z}_m$ is of first order in the $m$th-order  $x_3$-derivative of $\Tcm(x_3)$ at $x_3=0$.
\end{lem}

\noindent
{\it Proof of Lemma B.1.}
We solve Lyapunov-type equations (20) and (21) in \cite{TMC} to obtain ${\bf Z}_1$ and solve equations (23) to (25) in \cite{TMC} to obtain ${\bf Z}_m\  (m=2,3,\cdots, n)$.
By the chain rule of differentiation for the composite function $\Abb(x_3)=\Abb(x_3, \Tcm(x_3))=(a_{rs}(x_3))$, 
  ${d a_{rs}}/{d x_3}|_{x_3=0}$ is of first order in the first order $x_3$-derivative of $\Tcm(x_3)$ at $x_3=0$, whereas 
  $ {d^m a_{rs}}/{d x_3^m}|_{x_3=0}$ depends on  the  $x_3$-derivatives of $\Tcm(x_3)$ at $x_3=0$ up to those of order $m$ and is of first order in the $m$th-order $x_3$-derivative of $\Tcm(x_3)$ at $x_3=0$. 
Therefore,  (19) of \cite{TMC} implies that the right hand sides of (20) and (21) in \cite{TMC} are of first order in the first order $x_3$-derivative of $\Tcm(x_3)$ at $x_3=0$, whereas the right hand sides of (23) to (25) in \cite{TMC} depends on the  $x_3$-derivatives of $\Tcm(x_3)$ at $x_3=0$ up to those of order $m$, and are of first order in the $m$th-order $x_3$-derivative of $\Tcm(x_3)$ at $x_3=0$.
Hence the arguments  (26) through (27) in \cite{TMC} proves the lemma. \hfill $\Box$

\medskip

\noindent
{\it Proof of Theorem 3.1.}
The expression of $v_m$ in terms of ${\bf Z}_k\  (k=0,1,\cdots, m)$ was given by Section 6 of \cite{MNTW} for $m=1,2$ (see also (37) and (38) of \cite{TMC}).
These expressions were obtained from (\ref{nth-approximatesecular}) and
 the implicit function theorem, through which we  also have for a general $m$
\begin{equation}
  v_m =-\frac{ N_m }{m!\,D }
  \quad (m=1,2,\cdots, n),
 \label{v1vm}
\end{equation}
 where
 \begin{align}
D &=\frac{\partial R }{\partial v}\Bigr|_{v=v_0,\, \varepsilon=0}, \qquad
N_1 =\frac{\partial R }{\partial \varepsilon}\Bigr|_{v=v_0,\, \varepsilon=0}, 
\nonumber
\\[2mm]
 N_m &=\frac{\partial^m R }{\partial \varepsilon^m}\Bigr|_{\scriptstyle{v=v_0,}\atop{\!\!\!\!\scriptstyle{\varepsilon=0}}}
+\sum_{k+l=m, 
\atop{1\le l\le m,  \atop{ 0\le k\le m-1}}} 
\!\!\frac{m!}{k!\ l!}\ \frac{\partial^{k+l} R }{\partial \varepsilon^k\, \partial v^l}\cdot 
 \Bigl(\frac{d v_R }{d \varepsilon}\Bigr)^l
  \Bigr|_{\scriptstyle{v=v_0,}\atop{\!\!\!\!\scriptstyle{\varepsilon=0}}}
+\cdots
 \quad (m=2,3,\cdots, n)
\label{Nm}
\end{align}
and $\cdots$ on the right hand side of the preceding equation denotes a linear combination of the terms included in 
$$
\frac{\partial^{k+l} R }{\partial \varepsilon^k\, \partial v^l}\cdot 
\frac{d^{m-k-l}}{d \varepsilon^{m-k-l}}
  \Bigl(\frac{d v_R }{d \varepsilon}\Bigr)^l
  \Bigr|_{v=v_0,\, \varepsilon=0}
  \qquad (1<k+l<m, \ 1\le l).
$$
It then follows from 
(\ref{nth-approximatesecular}) that
$$
D =\frac{\partial R }{\partial v}\Bigr|_{v=v_0,\, \varepsilon=0}
=\frac{\partial }{\partial v}\det\,
{\bf Z}_0\,\Bigr|_{v=v_0},
$$
which does not depend on any component of the  $x_3$-derivatives of $\Tcm(x_3)$ at $x_3=0$, and it also follows that
\begin{equation}
\frac{\partial^m R }{\partial \varepsilon^m}\Bigr|_{\scriptstyle{v=v_0,}\atop{\!\!\!\!\scriptstyle{\varepsilon=0}}}
=
\frac{\partial^m }{\partial \varepsilon^m}\det\left[\,
{\bf Z}_0+{\bf Z}_1\, \varepsilon+{\bf Z}_2\, \varepsilon^2
+\cdots+ {\bf Z}_m\, \varepsilon^m\,
\right]\Bigr|_{\scriptstyle{v=v_0,}\atop{\!\!\!\!\scriptstyle{\varepsilon=0}}}
  \ \  (m=1,2,\cdots, n).
  \label{e-mth-derivative-R}
\end{equation}
Using the component-wise expression
$$
{\bf Z}_k=\bigl( Z^{(k)}_{ij} \bigr),
 \qquad k=0,1,2,\cdots, n,
$$
we observe from the definition of the determinant of a $3\times 3$ matrix that \begin{eqnarray*}
 \frac{\partial^m R }{\partial \varepsilon^m}\Bigr|_{\scriptstyle{v=v_0,}\atop{\!\!\!\!\scriptstyle{\varepsilon=0}}}
 &=&
  \frac{\partial^m }{\partial \varepsilon^m}\sum_{\sigma\in S_3}  \mbox{sgn}(\sigma) \sum_{j=0}^m Z^{(j)}_{1 \sigma(1)}\, \varepsilon^j 
 \cdot \sum_{k=0}^m Z^{(k)}_{2 \sigma(2)}\, \varepsilon^k
 \cdot  \sum_{l=0}^m Z^{(l)}_{3 \sigma(3)}\, \varepsilon^l\Bigr|_{\scriptstyle{v=v_0,}\atop{\!\!\!\!\scriptstyle{\varepsilon=0}}},
 \end{eqnarray*}
 where $S_3$ is the set of all  permutations of $\{1,2,3\}$.
 Hence,
\begin{eqnarray*}
 \frac{\partial^m R }{\partial \varepsilon^m}\Bigr|_{\scriptstyle{v=v_0,}\atop{\!\!\!\!\scriptstyle{\varepsilon=0}}}
 &=&  
  \sum_{\sigma\in S_3}  \mbox{sgn}(\sigma)  \, m!\,
  \left( Z^{(m)}_{1 \sigma(1)}Z^{(0)}_{2 \sigma(2)}Z^{(0)}_{3 \sigma(3)}
  + Z^{(0)}_{1 \sigma(1)}Z^{(m)}_{2 \sigma(2)} Z^{(0)}_{3 \sigma(3)}
  +Z^{(0)}_{1 \sigma(1)} Z^{(0)}_{2 \sigma(2)} Z^{(m)}_{3 \sigma(3)} \right)\Bigr|_{v=v_0}
   \\
  &&+\ 
  \sum_{\sigma\in S_3}  \mbox{sgn}(\sigma)  \, m!
  \!\!\!
\sum_{j+k+l=m, 
\atop{0\le j,k,l \le m-1}} \!\!\!  Z^{(j)}_{1 \sigma(1)}Z^{(k)}_{2 \sigma(2)}Z^{(l)}_{3 \sigma(3)}\Bigr|_{v=v_0},
 \end{eqnarray*}
 where the second term on the right hand side is neglected when $m=1$.  
 The first term is linear in  ${\bf Z}_m$, whereas the second term is nonlinear function of ${\bf Z}_k \ (k=0,1,\cdots, m-1)$.
 Moreover, any order of the partial derivative of $R(v, \varepsilon)$ with respect to $v$ does not affect the linearity in ${\bf Z}_m$.
 Hence only the first term on the right hand side of (\ref{Nm}) depends on (and is of first order in) ${\bf Z}_m$  and the remaining terms are  nonlinear functions of ${\bf Z}_k \ (k=0,1,\cdots, m-1)$.
This, combined with the preceding lemma and (\ref{v1vm}),  
proves the theorem.  \hfill $\Box$

\end{document}